\documentclass [prl,aps,amsmath,amssymb,showpacs,twocolumn,floats,epsf,superscriptaddress]{revtex4}
\usepackage{graphicx}
\usepackage{amsmath}
\usepackage{caption}
\usepackage{times}
\usepackage{amssymb}
\usepackage{bm}
\usepackage{color}

\begin{document}
\title{Interplay between electronic topology and crystal symmetry: Dislocation-line modes in topological band-insulators}
\author{Robert-Jan Slager}
\affiliation{Instituut-Lorentz for Theoretical Physics, Universiteit Leiden, P.O. Box 9506, 2300 RA Leiden, The Netherlands}
\author{Andrej Mesaros}
\affiliation{Department of Physics, Boston College, Chestnut Hill, MA 02467, USA}
\author{Vladimir Juri\v ci\' c}
\affiliation{Institute for Theoretical Physics, Utrecht University, Leuvenlaan 4, 3584 CE Utrecht, The Netherlands}
\author{Jan Zaanen}
\affiliation{Instituut-Lorentz for Theoretical Physics, Universiteit Leiden, P.O. Box 9506, 2300 RA Leiden, The Netherlands}

\begin{abstract}
 We elucidate the general rule governing the response of dislocation lines in three-dimensional topological band insulators. According to this ${\bf K}\text{-}{\bf b}\text{-}{\bf t}$ rule, the lattice topology, represented by dislocation lines oriented in direction ${\bf t}$ with Burgers vector ${\bf b}$,  combines with the electronic-band topology, characterized by the band-inversion momentum ${\bf K}_{\rm inv}$, to produce gapless propagating modes when the plane orthogonal to the dislocation line features a band inversion with a nontrivial ensuing flux $\Phi={\bf K}_{\rm inv}\cdot {\bf b}\,\, ({\rm mod\,\,2\pi})$. Although it has already been discovered by Y. Ran {\it et al.}, Nature Phys. {\bf 5}, 298 (2009),  that dislocation lines host propagating modes, the exact mechanism of their appearance in conjunction with the crystal symmetries of a topological state is provided by the ${\bf K}\text{-}{\bf b}\text{-}{\bf t}$ rule . Finally, we discuss possible experimentally consequential examples in which the modes are oblivious for the direction of propagation, such as the recently proposed topologically-insulating state in electron-doped BaBiO$_3$.
\end{abstract}


\maketitle

Topological band-insulators (TBIs) represent a new class of quantum materials that, due to the presence of time-reversal symmetry (TRS), feature an insulating bulk bandgap together with metallic edge or surface states protected by a ${\mathbb Z}_2$ topological invariant\cite{Kane-Mele-PRL2005,fukane2006,moore2007,fukane2007a}. This  ${\mathbb Z}_2$ classification of topological insulators is a part of the classification of free gapped fermion matter in the presence of the fundamental antiunitary time-reversal and particle-hole symmetries, the so-called tenfold way \cite{schnyder2008,schnyderNJP2010, kitaev2009}. On the other hand, topologically-insulating crystals break continuous translational and  rotational symmetries down to discrete symmetries mathematically characterized by the space groups.
By considering the crystal symmetries, an extra layer in this ${\mathbb Z}_2$ classification of TBIs has been recently uncovered \cite{Nat-Phys-us2013}.  This space group classification of TBIs results in the enrichment of the tenfold way with extra phases, such as  crystalline (or ``valley'') phases \cite{Fu2011} and translationally-active phases, the latter featuring an odd number of band-inversions at non-$\Gamma$ points in the Brillouin zone (BZ)  \cite{Nat-Phys-us2013,us-prl2012}.
Dislocations are of central interest in this endeavor, being the topological defects exclusively related to the lattice translations.
In two dimensions (2D), the role of these lattice defects has been recently elucidated in TBIs \cite{us-prl2012, Dunhaiarxiv2013}, as well as in topological superconductors \cite{asahinagaosa2012,hughesyaoqi2013} and interacting topological states \cite{barkeshli-qiprx2012,barkeshli-qi-prb2013,andrejran2013}. In particular, it has been shown that these lattice defects in two-dimensional TBIs act as probes of distinct topological states through binding of the localized zero-energy modes in their core \cite{us-prl2012}.

\begin{figure*}
 \includegraphics[scale=0.55]{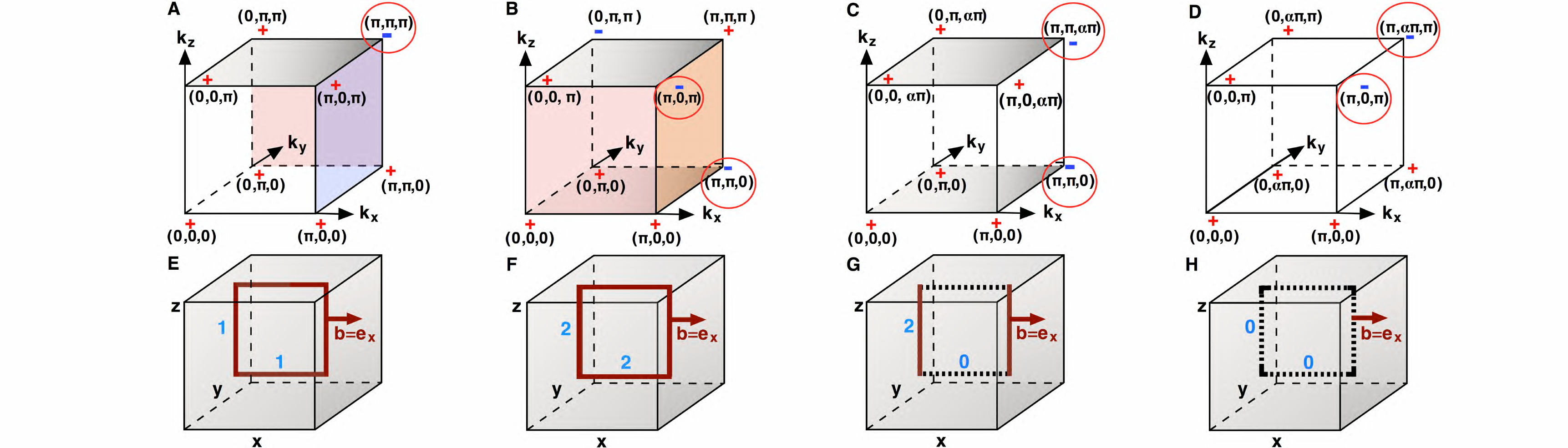}
\caption{(Color online) Illustration of the ${\bf K}\text{-}{\bf b}\text{-}{\bf t}$ rule relating the electronic topology in the momentum space,  and the effect of dislocations in real space. Panels A to D show the electronic-band topology of the $R$  [$T\text{-}p3(4)_{R}$] and $M$ [$T\text{-}p3(4)_{M}$] phases on a simple cubic lattice and the $M-R$  ($p4_{M,R}$) and $X'-R$ ($p4_{X',R}$) weak phases on tetragonal lattices. A dislocation with Burgers vector ${\mathbf b} = {\bf e}_x$  acts on the encircled TRI
momenta in the planes orthogonal to the dislocation line. As a result, the colored planes host an effective $\pi$ flux. The resulting number of Kramers pairs of helical modes along the edge and screw parts of the loop is indicated with the blue number.
(A) The symmetric $R$ phase has a topologically nontrivial plane hosting a $\pi$ flux orthogonal to any of the three crystallographic directions and hence any dislocation loop binds modes along the entire core, as shown for a loop in the $\hat{x}-\hat{z}$ plane, panel E. (B) In the $M$ phase, translationally active phases in the TRI planes orthogonal to $k_z$ and a crystalline phase in $k_x=\pi$ plane host $\pi$ fluxes. Hence the dislocation loop binds two pairs of modes, as displayed in panel F.
These modes are symmetry-protected against mixing. (C) In the $M-R$ phase, only the TRI planes normal to $k_{z}$ host an effective $\pi$-flux and hence the same dislocation loop binds modes only to the edge-dislocation parts, as displayed in panel G. These modes are not protected against mixing. (D) In the $X'-R$ phase all TRI planes orthogonal to the dislocations lines have a trivial flux, and, according to the ${\bf K}\text{-}{\bf b}\text{-}{\bf t}$ rule,  neither the edge nor the screw dislocation of the loop binds  modes, as illustrated in panel H.
}
\label{figureone}
 \end{figure*}

Although early on it was identified that in three-dimensional TBIs dislocation lines support propagating helical modes \cite{ranNatPhys2009}, the precise role of dislocations have not been explored thoroughly \cite{teokane2010,ranarxiv2010,imuraprb2011}. In particular, the relation between the lattice symmetry and the electronic topology, as well as the  characterization of these topological states through the response of the dislocation lines has not been addressed. Dislocations in three dimensions (3D) are defects with richer structure than their two-dimensional counterparts. They form lines ${\bf l}(\tau)$  (parametrized by $\tau\in[0,1]$), characterized by a tangent vector ${\bf t}\equiv d{\bf l}/d\tau$, with the discontinuity introduced to the crystalline order described by a Burgers vector ${\bf b}$. Both these vectors can only be oriented along the principal axes of the crystal, and a screw (edge) dislocation is obtained when ${\bf b}\parallel {\bf t}$ (${\bf b}\perp {\bf t}$), see Fig.1. Moreover, the Burgers vectors are additive, and in full generality we can thus consider only dislocations with the Burgers vectors equal to Bravais lattice vectors. The crucial observation is that translational lattice symmetry is preserved along the dislocation line. Therefore, the full lattice Hamiltonian in the presence of a dislocation oriented along the $z$-axis (${\bf t}={\bf e}_z$) can be written as
\begin{equation}\label{eq:ham-dim-red}
H_{\rm 3D}(x,y,z)=\sum_{k_{z}}e^{ik_z z}H_{\rm eff}^{\rm 2D}(x,y,k_z).
\end{equation}
Notice that the above 2D lattice Hamiltonian possesses the (wallpaper group) symmetry of the crystallographic plane orthogonal to the dislocation line, because the Burgers vector is a Bravais lattice vector.

On the other hand, the electronic topology of a TBI is characterized by the band-inversions at time-reversal invariant (TRI) momenta  ${\bf K}_{\rm inv}$ in the BZ \cite{Nat-Phys-us2013}. A dislocation disturbs the crystalline order only microscopically close to its core. We can therefore use elastic continuum theory to describe its effect at low energies. The elastic deformation of the continuous medium is encoded by a distortion field ${\bm\varepsilon}_i$ of the global Cartesian reference frame ${\bf e}_i$, $e_i^\alpha=\delta_i^\alpha$, with $i,\alpha=1,2,3$ \cite{Kleinert,Turski}. The  momentum near the transition from a topologically trivial to a nontrivial phase with the bandgap closing at the momentum  ${\bf K}_{\rm inv}$ is $k_i=({\bf e}_i+{\bm\varepsilon}_i)\cdot({\bf K}_{\rm inv}-{\bf q})$, with $q\ll K_{\rm inv}\sim 1/a$ as the momentum of the low-energy electronic excitations, ${\varepsilon}\sim a/r$, $a$ is the lattice constant, and $r$ is the distance from the defect core. Therefore the  dislocation gives rise to a $U(1)$ gauge field $A_i=-{\bm\varepsilon}_i\cdot{\bf K}_{\rm inv}$ that minimally couples to the electronic excitations, ${\bf q}\rightarrow{\bf q}+{\bf A}$.
The translational symmetry then implies that the gauge field has nontrivial components only in the plane orthogonal to the dislocation line, consistent with Eq.\ (\ref{eq:ham-dim-red}), carrying an effective flux $\Phi={\bf K}_{\rm inv}\cdot {\bf b}$, as we demonstrate below.

Consider an edge and a screw dislocation both oriented  along the $z$-axis. We use that for the edge dislocation with Burgers vector ${\bf b}=a {\bf e}_x$, the dual basis in the tangent space at the point ${\bf r}$ is ${\bf E}^x=\left(1-\frac{a y}{2\pi r^2 }\right){\bf e}^x + \frac{a x}{2\pi r^2}{\bf e}^y,\, {\bf E}^y={\bf e}^y,\,{\bf E}^z={\bf e}^z$, while for the screw dislocation with ${\bf b}=a {\bf e}_z$, we have ${\bf E}^x={\bf e}^x ,\, {\bf E}^y={\bf e}^y,\,{\bf E}^z=\frac{b}{2\pi r^2}(-y{\bf e}^x+x{\bf e}^y)+{\bf e}^z$; $r^2\equiv x^2+y^2$ \cite{Kleinert}. The corresponding distortion for both an edge and a screw dislocation is then readily obtained to leading order in $a/r$ to be $\bm{\varepsilon}_x=\frac{y}{2\pi r^2}{\bf b},\,\, {\bm\varepsilon}_y=-\frac{x}{2\pi r^2}{\bf b}$ (see Supplemental Material in \cite{Supplementary} for details), and the corresponding gauge potential ${\bf A}({\bf r})$  has nontrivial components in the plane orthogonal to the dislocation line, ${\bf A}\cdot {\bf t}=0$, and
\begin{equation}\label{eq:flux}
{\bf A}=\frac{-y {\bf e}_x+x{\bf e}_y}{2\pi r^2}({\bf K}_{\rm inv}\cdot {\bf b})\equiv\frac{-y {\bf e}_x+x{\bf e}_y}{2\pi r^2}\Phi.
\end{equation}
When this flux  $\Phi\,({\rm mod }\, 2\pi)$ is nonzero, the dislocations  carry propagating helical modes provided that  the 2D Hamiltonian in a TRI plane orthogonal to the dislocation line,  $\hat{\bf t}\equiv{\bf t}/|{\bf t}|$, hosts  a band inversion.
The lattice symmetry that relates the band-inversion momenta then protects these modes. This is what we refer to as the ${\bf K}\text{-}{\bf b}\text{-}{\bf t}$ rule.
This rule and the following descendant construction together imply  that the bound states for a given ${\bf k}\cdot\hat{\bf t} $ momentum combine into the spectrum
of the propagating helical modes along the dislocation line. For ${\bf k}\cdot\hat{\bf t}={\bf K}_{\rm inv}\cdot\hat{\bf t}$, the system develops a Kramers pair of true {\it zero} modes $\Psi_0\equiv(\psi_0,T\psi_0)^\top$, with $T$ as the time-reversal operator and $T^2=-1$ \cite{us-prl2012}. This is a consequence of the fact that by definition the ${\bf K}_{\rm inv}$ point is hosting a band inversion. Henceforth, the physics is essentially captured by a Dirac Hamiltonian with negative mass.
Deviating from this 'parent'  momentum by $ {\bf q}\cdot \hat{\bf t}$,  the effective low-energy Hamiltonian for the propagating modes then generally develops a linear gap $H_{\rm eff}\sim {\bf q}\cdot \hat{\bf t}$, to lowest order. The gapped Kramers pair of descendant states is then present as long as $H_{\rm eff}({\bf q})$ remains topologically nontrivial,  and may then be captured by $H_{\rm eff}=v_{ t}\Sigma_3({\bf q}\cdot \hat{\bf t})+{\mathcal O}(q^2)$. Here,  the Pauli matrix $\Sigma_3$ acts in the two-dimensional Hilbert space of the dislocation modes, which are of the form $\Psi_{{q_t}}\equiv(\psi_0\,e^{i ({\bf q}\cdot\hat{\bf t})({\bf r}\cdot\hat{\bf t})},(T\psi_0)\,e^{-i ({\bf q}\cdot\hat{\bf t})({\bf r}\cdot\hat{\bf t})})^\top$, and $v_{ t}$ is the characteristic velocity, which is set by the symmetries and details of the band structure.

We thus see how the dislocation line ${\bf t}$, the Burgers vector ${\bf b}$, and the TRI momenta  ${\bf K}_{\rm inv}$, through the ${\bf K}\text{-}{\bf b}\text{-}{\bf t}$ rule, conspire into a precise condition determining the existence of the dislocation propagating modes in certain directions in a topologically-insulating phase, consistent with the space group classification \cite{Nat-Phys-us2013} (Fig. 1). By varying ${\bf t}$ and ${\bf b}$ in all directions, the number of  "parent" zero modes in any projection plane is in one-to-one correspondence with the space group classification in terms of the ${\bf K}_{\rm inv}$ momenta. Specifically, for a (translationally-active) phase with a single ${\bf K}_{\rm inv}$ momentum, edge and screw dislocations bind a single Kramers pair of helical modes if ${\bf K}_{\rm inv}\cdot{\bf b}\neq0$.
 In case of a (translationally-active or a crystalline) phase with multiple ${\bf K}_{\rm inv}$ momenta, a projected 2D system may also result in a double pair of modes if the effective system entails a 2D crystalline phase. The two pairs are then protected by the symmetries relating the ${\bf K}_{\rm inv}$ momenta.
Most interestingly, there is the possibility that both edge and screw dislocations bind modes in any crystal direction resulting in propagating modes along the full dislocation loops. In particular, in a completely isotropic lattice with $O_h$ crystal symmetry a strong variant of this effect can be realized: the gapless states in the dislocation loop that propagate in the way completely oblivious to the lattice directions.
 In contrast,  in a weak phase characterized by the weak topological invariant  ${\bf M}$, protected helical modes are  obtained only when  the product ${\bf M}\cdot {\bf b}\,\, ({\rm mod}\, 2\pi)$ is nontrivial \cite{ranNatPhys2009,teokane2010,ranarxiv2010,moore2007}. This is consistent with the fact that double pairs of modes originating from two  ${\bf K}_{\rm inv}$ not related by symmetry may be gapped out. However, according to the ${\bf K}\text{-}{\bf b}\text{-}{\bf t}$ rule if the ${\bf K}_{\rm inv}$ are related by symmetry such pairs are protected from gapping out.  We expect that the stability of the dislocation modes in 3D TBIs against a  time-reversal invariant disorder is in one-to-one correspondence with the stability of the 2D topological phase yielding the helical modes. Therefore, the most stable dislocation modes should originate from a 2D $\Gamma-$phase, followed by the ones   arising from translationally-active and crystalline phases \cite{Nat-Phys-us2013}.

\begin{figure*}
 \includegraphics[scale=0.45]{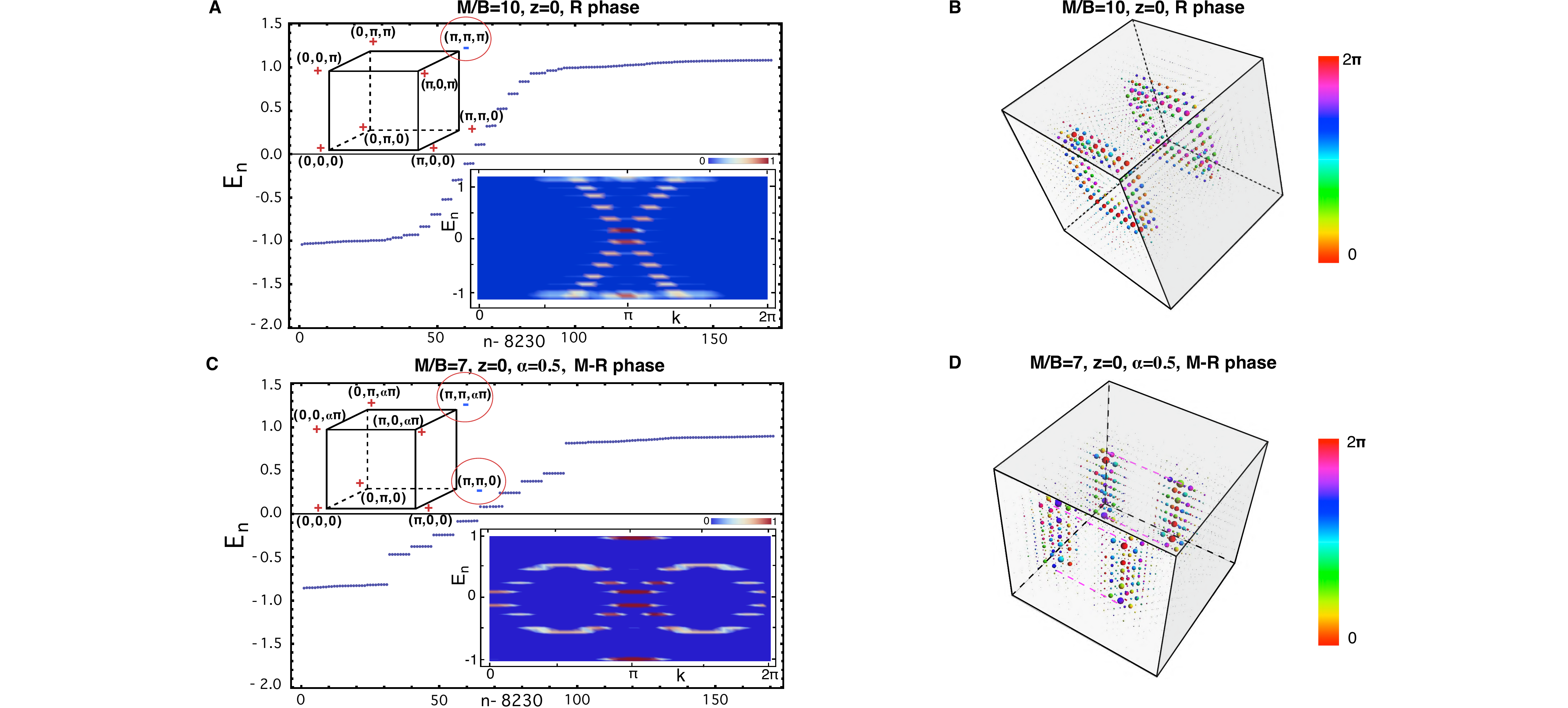}
\caption{ (Color online) The effect of dislocation lines in 3D topological band insulators. Edge and screw dislocations with Burgers vector $\mathbf{b}=\mathbf{e}_{x}$ are treated simultaneously by considering dislocation loops of $8\times8$ sites within the tight-binding model \eqref{eq::Hamtbmain} on the lattice with  $16\times16\times16$ sites in case of periodic boundary conditions, see \cite{Supplementary} for details. Panels A and C show the spectrum of the dislocation modes traversing the gap in case of the cubic $R$ phase [$T\text{-}p3(4)_{R}$] and tetragonal $M-R$ ($p4_{M,R}$) phases, with the corresponding electronic topological configurations shown in the insets on the upper left. The circles indicate the TRI momenta hosting band-inversions that yield effective $\pi$-fluxes. Additionally, the spectral density as function of the momentum $\mathbf{k}$ along the dislocation line is displayed in the insets on the lower right. We find a single cone for the $R$ phase, and a double cone for the  $M-R$ phase, consistent with the  ${\bf K}\text{-}{\bf b}\text{-}{\bf t}$ rule. The energy levels are now eightfold degenerate as both  $k_{z}=0$ and $k_{z}=\pi$ planes are topologically nontrivial, and thus each yields propagating modes due to the effective $\pi$ flux introduced by the dislocations. The zero modes are offset due to a finite system size (see also Fig.\ S11 in Ref.\ \onlinecite{Supplementary}). Panels B and D display the real space localization of the modes, where the weight of the wavefunction is indicated by the size of the circles and the color coding indicates the corresponding phase. Most importantly,  the cubic $R$ phase features   topologically-protected propagating dislocation modes along the complete loop. In contrast, the  $M-R$ phase binds modes only along the edge-dislocation parts. In particular, we find eight energy levels in correspondence with  the number of allowed momenta along the dislocation. }
\label{figuretwo}
 \end{figure*}

These general statements can be illustrated by a simple tight-binding model with two orbitals with different parity and two spin states \cite{ZhangNatPhys2009} on the simple cubic lattice with space group $pm\bar{3}m$
\begin{equation}\label{eq::Hamtbmain}
H_{nn}=A(\gamma_1\sin k_x +\gamma_2 \sin k_y+\gamma_3\sin k_z)+M_{nn}\gamma_{0},
\end{equation}
with ${\bf k}$ as the electron momentum, the $\gamma$-matrices acting in the orbital and the spin spaces, and we use natural units ($\hbar=c=a=1$).
Topological phases of the model \eqref{eq::Hamtbmain} can be tuned with the mass term $M_{nn}=m-2B(3-\cos k_x-\cos k_y-\cos k_z )$. The parameters $A$($B$) are related to the inter-(intra-)orbital hoppings, and $m$ is related to the difference in the onsite energies for different orbitals \cite{Supplementary}.
Let us assume that the system is in an $R$-phase (labeled $T\text{-}p3(4)_{R}$ in Ref.~\onlinecite{Nat-Phys-us2013}), which is characterized by a band-inversion at the $R$ point [momentum $(\pi,\pi,\pi)$] in the BZ,  (Figs. 1A and 2). If $\mathbf{b}=\mathbf{e}_{x}$ and ${\bf t}=\mathbf{e}_{z}$, the effective Hamiltonian in the $k_{z}=\pi$ plane reduces to the $\pi$ flux problem in the two-dimensional $M$-phase  (labeled  $T-p4_{M}$ in Ref.~\onlinecite{Nat-Phys-us2013}), which is defined by the band-inversion at the $M$ point, and hence possesses zero modes.
  For the family of Hamiltonians \eqref{eq:ham-dim-red} then any momentum $k_z=\pi+q_z$ infinitesimally close to $k_z=\pi$ yields the would-be zero modes in the absence of the term $\sim \gamma_3$. When included, this term gives rise to the linearly dispersing propagating modes along the dislocation line, according to the above $\mathbf{K}\text{-}\mathbf{b}\text{-}{\mathbf t}$ rule.
  Furthermore, the same rule implies that an edge or a screw dislocation along any Bravais lattice vector in the simple cubic lattice effectively acts as a $\pi$ flux.
Hence, when these defects are joined in a loop in the $x-z$-plane with $\mathbf{b}=\mathbf{e}_{x}$, we expect propagating modes, which are indeed found in our numerical computations. Clearly, TRS together with the crystal symmetry protects the modes from backscattering.

Next we consider the system in  the three-dimensional $M$-phase [$T-p3(4)_{M}$], characterized by band-inversions located at the momenta $M\equiv(\pi,\pi,0)$, $X'\equiv(\pi,0,\pi)$, $Y'\equiv(0,\pi,\pi)$, which are related by the threefold rotational symmetry (Figs. 1B, and S5 and S10 in the Supplemental Material \cite{Supplementary}). According to the ${{\bf K}\text{-}{\bf b}}\text{-}{\bf  t}$ rule, for the edge-segment of the dislocation loop, $k_z=0$ and $k_z=\pi$ planes both host a $\pi$ flux, originating from the $M$ and $X'$ points. Additionally, the $k_x=\pi$ plane hosts  a two-dimensional crystalline phase, and thus the screw-dislocation parts also host two pairs of modes. Therefore, there is a total of two Kramers pairs of gapless dislocation modes along the loop, which are {\it protected by symmetry}. Note that their existence crucially depends on the fact that the $M$ and $X'$ momenta are symmetry-related. Were this not the case, the weak index of this $T-p3(4)_{M}$ phase, $M_i=(0,0,0)$, would predict no dislocation modes at all.

In contrast, let us now break the cubic symmetry by considering the $M-R$-phase  ($p4_{M,R}$) on a tetragonal lattice with $a_x=a_y=a$, $a_z=a/\alpha$, where $a_i$ is the lattice constant in the direction ${\bf e}_i$, and $\alpha\neq1$ is the lattice deformation parameter (Figs. 1C and 2). Notice here also a subtle difference between such a weak phase and a crystalline (valley) phase, which has an even number of band-inversions {\it protected} by a 3D space group symmetry. The usual strong and weak indices  \cite{moore2007,fukane2007a,ranNatPhys2009} in this phase are $(\nu; M_i)=(0;0,0,1)$, and thus  $\mathbf{M}\cdot\mathbf{b}=0$. Nonetheless,  the $k_z=0$ and $k_z=\alpha\pi$ planes are topologically nontrivial therein (Fig. 1 and \cite{Supplementary}). As a result, for a dislocation loop with ${\bf b}={\bf e}_x$, according to the ${{\bf K}\text{-}{\bf b}}\text{-}{\bf t}$ rule, we find modes bound only to the edge-dislocation parts.
As both these planes contribute the midgap states, we expect a double Kramers pair of the propagating metallic states, which our numerical computations indeed confirm. However, these modes can mix in the dislocation loop, since no symmetry relates the momenta $M$ and $R$ giving rise to them.

Finally, we consider the tetragonal  $X'-R$ ($Y'-R$) phase obtained by deforming the cubic lattice in the ${\bf e}_y$ (${\bf e}_x$) direction with the corresponding band-inversions at $(\pi,0,\pi)$ $[(0,\pi,\pi)]$ and  $(\pi,\alpha\pi,\pi)$ $[(\alpha\pi,\pi,\pi)]$ momenta (Fig.\ 1D).  No dislocation modes appear in the {\bf $X'-R$} phase, whereas in the {\bf $Y'-R$} phase only the band-inversion at momentum $(\alpha\pi,\pi,\pi)$ contributes a $\pi$ flux, thus yielding the modes for both types of dislocations.
These results are therefore similar as in  the ${R}$-phase on the cubic lattice, as also confirmed by numerical computations showing modes propagating along the entire loop in the $x-z$ plane. However, the difference is that not all directions are equivalent here, and hence velocities of the modes along the loop are anisotropic.

The response of dislocations in TBIs is experimentally consequential as these defects are ubiquitous in any crystals. The electron-doped mixed-valent perovskite oxide $\text{BaBiO}_{3}$ with a simple cubic crystal symmetry has been recently predicted to be a TBI  \cite{BinghaiNatphys2013}. More interestingly, the topological phase is believed to be the symmetric $R$-phase [$T\text{-}p3(4)_{R}$], which should host propagating gapless modes along any dislocation loop, as we have found here. The effective tight-binding model \eqref{eq::Hamtbmain} describes this phase with the parameters $A=2.5$eV$\cdot$\AA, $B=9.0$eV$\cdot$\AA$^2$, $M=5.08$eV, and the lattice constant $a=4.35$\AA~  \cite{BinghaiNatphys2013}, implying that the velocity of the dislocation modes is $v=A/\hbar=4.3\times10^5$m/s, while the localization length  $\lambda\sim \sqrt{B/(12Ba^{-2}-M)}\simeq4.8$\AA~ \cite{NPB-us}. The enhanced density of states near the dislocation line in a non-$\Gamma$ TBI \cite{hsieh2008,ando2012,xu2012,story2012} should be observable by local probes, such as scanning tunnelling microscopy,  where the dislocation reaches the crystal surface. The surface states hybridize with the dislocation modes, and the precise local redistribution of the increased density of states can be modeled for the specific experimental situation using our theory. Angle-resolved photoemission spectroscopy may also be useful for mapping out these states, since the surface irregularities should not affect this probe.

We hope that our findings will motivate further studies of the role of dislocations in TBIs with different crystal symmetries, and especially their knotting which should play the role of braiding for these topological defects and might be relevant for quantum computation. Finally, our results based on the ``intuitive'' classification of the dislocations through the Burgers vectors ultimately call for a mathematically precise characterization of these defects in terms of the homotopy classes for the discrete space groups.

 {\it Note added.} Recently, we became aware of the work by Shiozaki and Sato \cite{shiozaki-sato}, where the response of dislocations in 3D TBIs was studied using $K-$theory.

 This work is part of the D-ITP consortium, a program of the Netherlands Organisation for Scientific Research (NWO) that is funded by the Dutch Ministry of Education, Culture and Science (OCW). This work is supported by the Dutch Foundation for Fundamental Research on Matter (FOM).  V.\ J.\ acknowledges the support of NWO. The authors acknowledge fruitful discussions with Xiao-Liang Qi and Zhi-Xun Shen.

\clearpage
\newpage
\begin{widetext}
\begin{center}
{\LARGE \underline{Supplementary Material:}}\\
{\LARGE Interplay between electronic topology and crystal symmetry: Dislocation-line modes in topological band-insulators}
\end{center}
\par
\begin{center}
\vspace{1em}
{\large Robert-Jan Slager, Andrej Mesaros, Vladimir Juri\v ci\' c and Jan Zaanen}
\end{center}
\section*{A. Model details}
 We here present the details of the model that we use to obtain topological states on cubic and tetragonal lattices. The tight-binding model
 contains two spin degenerate  orbitals with a Hamiltonian
\begin{equation*}\label{eq::general}
H=\epsilon(\mathbf{k})\mathbf{1}+\sum_{\alpha}{d}_{\alpha}(\mathbf{k})\gamma_{\alpha}+\sum_{\alpha\beta}{d}_{\alpha\beta}\gamma_{\alpha\beta}, \tag{S1}
\end{equation*}
where $\gamma_{\alpha}$ are the five Dirac matrices obeying the Clifford algebra $\{\gamma_\alpha,\gamma_\beta\}=2\delta_{\alpha\beta}$, $\gamma_{\alpha\beta}$ are the ten commutators $\gamma_{\alpha\beta}=\frac{1}{2i}[\gamma_{\alpha},\gamma_{\beta}]$.
Specifically, we take the following basis of the $\gamma$-matrices
\begin{equation}\nonumber
\gamma_{0}=\sigma_{0}\otimes\tau_{3}\qquad \gamma_{1}=\sigma_{1}\otimes\tau_{1}, \qquad \gamma_{2}=\sigma_{2}\otimes\tau_{1}, \qquad\gamma_{3}=\sigma_{3}\otimes\tau_{1},\qquad\gamma_{5}\equiv - \gamma_0\gamma_1\gamma_2\gamma_3=\sigma_{0}\otimes\tau_{2},
\end{equation}
with ${\bm \sigma}$ and ${\bm \tau}$ being the standard Pauli matrices acting in the spin and orbital space, respectively; $\mathbf{1}=\sigma_0\otimes\tau_0$ with $\sigma_0,\tau_0$ as the $2\times2$ unity matrices.

The explicit form of the Hamiltonian \eqref{eq::general} is then determined by exploiting the symmetries. For the tight-binding model we assume a well-defined parity, in
 addition to time-reversal symmetry. Time-reversal symmetry is represented by the operator $T=\vartheta K$, with $\vartheta=i\sigma_{2}\otimes\tau_{0}$ and $K$ as complex conjugation, while the parity operator is $P=\gamma_{0}$. As the commutators $\gamma_{\alpha\beta}$ transform under time-reversal and parity with the opposite sign, the assumption of having both these symmetries thus results in ${d}_{\alpha\beta}=0$ in Eq.\ \eqref{eq::general}.
The remaining functions ${d}_{\alpha}({\bf k})$ of the effective model can then be obtained from the theory of invariants. Let us consider the
 space group $pm\bar{3}m$ with the point group $O_h$.  This point group has two one-dimensional, one two-dimensional and two three-dimensional irreducible representations.
 The matrix $\gamma_{0}$ anticommutes with the $\gamma$-matrices  $\{\gamma_{1}, \gamma_{2}, \gamma_{3}\}$  and therefore represents the mass term, which is by construction even under the parity. The corresponding mass term has to be rotationally-symmetric and therefore an even polynomial in the momentum ${\bf k}$. Using also that the $\gamma$-matrices  $\{\gamma_{1}, \gamma_{2}, \gamma_{3}\}$ form a three-dimensional representation under rotations in $O_h$, we arrive at the following minimal continuum Hamiltonian that captures topologically nontrivial phases
\begin{equation*}\label{eq::Hamcon}
H_{eff}= A(k_x\gamma_{1}+k_y\gamma_{2}+k_z\gamma_{3})+M_{nn}\gamma_{0}+\mathcal{O} (k^3),\tag{S2}
\end{equation*}
with $M_{nn}=m-2B(k_{x}^2+k_{y}^2+k_{z}^2 )$ and we dropped the term $\sim {\mathbf 1}$ not relevant for the topological analysis here. This Hamiltonian generalizes the 2D Bernevig-Hughes-Zhang (BHZ) to three dimensions (3D) and its different incarnations have already been used to describe topological states in Bismuth-based compounds [S1].

Let us subsequently determine the lattice-regularized version of the effective model \eqref{eq::Hamcon}. Considering a simple cubic lattice with nearest-neighbor (nn) hoppings, we obtain the following tight-binding Hamiltonian
\begin{equation*}\label{eq::Hamtb}
H_{nn}= A (\gamma_{1}\sin k_x+\gamma_{2}\sin k_y+ \gamma_{3}\sin k_z)+M_{nn}\gamma_{0}, \tag{S3}
\end{equation*}
with the effective mass parameter
\begin{equation}\nonumber
 M_{nn}=m-2B(3-\cos k_{x} -\cos k_{y} -\cos k_{z} ).
\end{equation}
We set the lattice constant $a=1$, and the parameters $A$ ($B$) represent hopping amplitudes between different (same) orbitals, while $m$ is the difference of the onsite energies between the two orbitals, similar as in the two-dimensional (2D) BHZ model. The hoppings in the Hamiltonian (\ref{eq::Hamtb}) are equal in the three orthogonal directions due to the rotational symmetry.
The above Hamiltonian reduces to the continuum Hamiltonian \eqref{eq::Hamcon} when expanded around the $\Gamma$ or the $R$-point located at the momentum $(\pi,\pi,\pi)$ in the Brillouin zone (BZ). Additionally, in order to enrich the phase diagram, we add next-nearest-neighbor hopping terms in the exact same fashion as the nearest-neighbor ones. Notice that on a cube each site has four next-nearest in each of the three mutually orthogonal crystallographic planes resulting in the Hamiltonian
\begin{align}
H_{nnn}&=\frac{\tilde{A}}{2}[\sin(k_{x}+k_{y})(\gamma_{1}+\gamma_{2})+\sin(-k_{x}+k_{y})(-\gamma_{1}+\gamma_{2})]&\nonumber\\
&+\frac{\tilde{A}}{2}[\sin(k_{x}+k_{z})(\gamma_{1}+\gamma_{3})+\sin(-k_{x}+k_{z})(-\gamma_{1}+\gamma_{3})]&\nonumber\\
&+\frac{\tilde{A}}{2}[\sin(k_{y}+k_{z})(\gamma_{2}+\gamma_{3})+\sin(-k_{y}+k_{z})(-\gamma_{2}+\gamma_{3})]&\nonumber\\
&-4\tilde{B}[3-\cos(k_{x})\cos(k_{y})-\cos(k_{x})\cos(k_{z})-\cos(k_{y})\cos(k_{z})]\gamma_{5},&\tag{S5}
\end{align}
so the total Hamiltonian of the tight-binding model on the cubic lattice is
\begin{equation*}\label{eq::tb}
H_{\rm TB}=H_{nn}+H_{nnn}\tag{S6}
\end{equation*}
Different topological phases are obtained by varying the mass term multiplying $\gamma_{5}$ matrix. The topological phase transitions occur when the mass term vanishes at the time-reversal invariant momenta in the BZ and the corresponding phases can be characterized by the mass term at these special momenta [S2].  As a result the phase diagram, with $A=1$,  is readily obtained as a function of $M/B$ and $\tilde{B}/B$, see Fig. S1.

By lowering the full rotational symmetry of the Hamiltonian (\ref{eq::Hamtb}), one obtains a minimal tight-binding Hamiltonian describing topological phases
on a tetragonal lattice
\begin{align}\label{eq:tetragonal}
H^{\rm tetragonal}_{\rm nn}&=A_x\gamma_1\sin k_x +A_y\gamma_2\sin k_y +A_z\gamma_3\sin k_z &\nonumber\\
&+[m-2B_x(1-\cos k_x)-2B_y(1-\cos k_y)-2B_z(1-\cos k_z)]\gamma_{5},&  \tag{S7}
\end{align}
which we will use later.

\begin{figure}[h]
 \includegraphics[scale=0.8432]{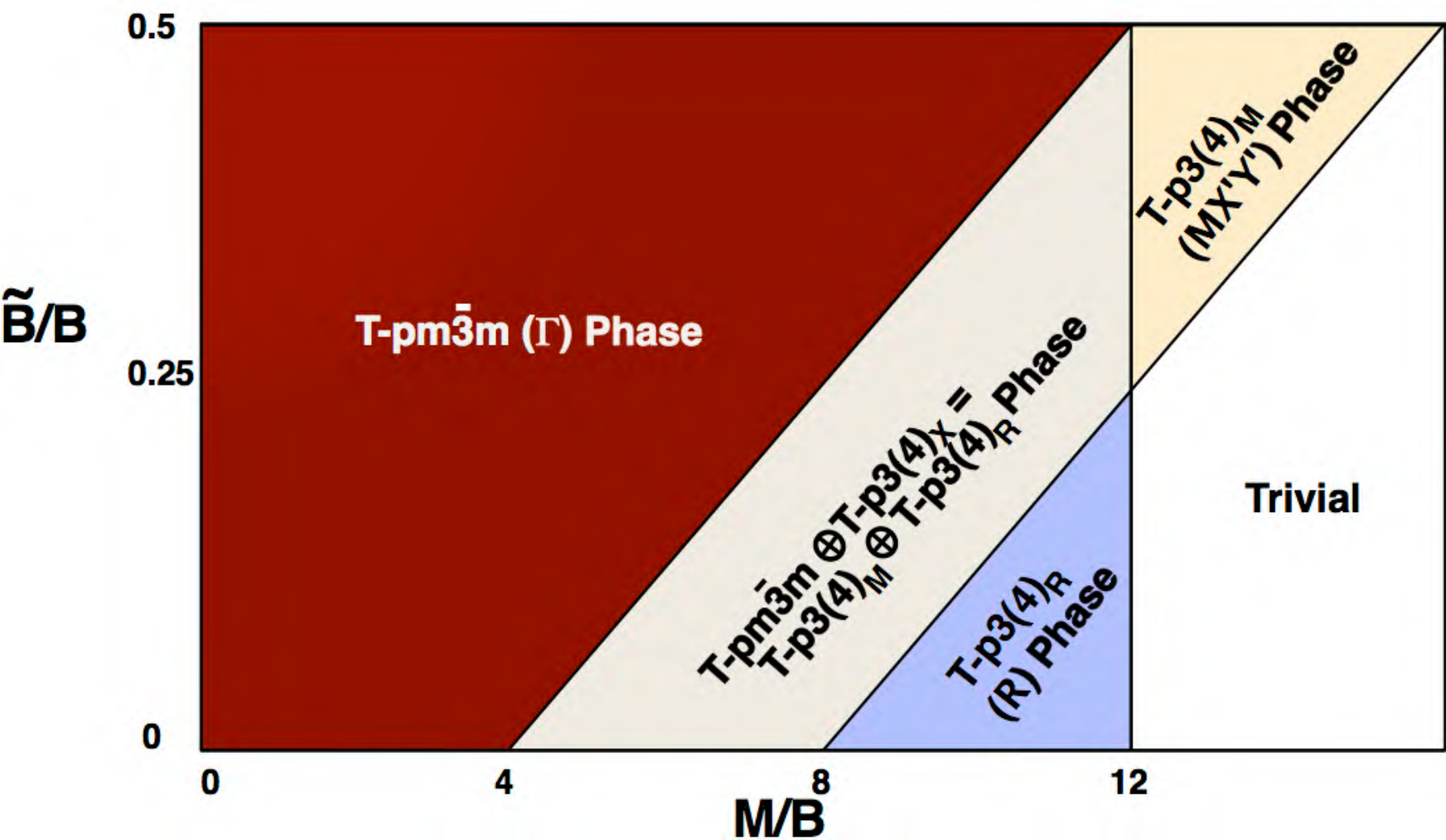}
 \caption{\label{Fig. S1}Figure\ S1: The phase diagram of the tight-binding model \eqref{eq::tb}, with space group $pm\bar{3}m$ and $O_{h}$ point-group symmetry. As function of the  parameter $M/B$ and the nearest neighbor hopping $\tilde{B}/B$ the mass parameter at the time-reversal invariant momenta can be tuned to produce the different electronic topological configurations as shown in the figure.  }
  \end{figure}

\clearpage
\newpage

\section*{B. Analytical and numerical description of the dislocation modes in three-dimensional topological band-insulators }
Let us now consider the effect of dislocations in the coarse-grained continuum theory obtained from the tight-binding model introduced in the previous section. Such lattice defects are described  within the elastic continuum theory using vielbeins, encoding the map from the perfect lattice to the distorted lattice [S3]. The torsion $T^{i}$ and curvature $R^{i}_{j}$ are then related to the vielbeins $E^i_\alpha$ and the spin connection $\omega^i_j$ by the Einstein-Cartan structure equations
\begin{align*}\label{eq::Einstein-Cartan}
&T^{i}=dE^{i}+\omega^{i}_{j}\wedge E^{j} \\
&R^{i}_{j}=d\omega^{i} _{j}+\omega^{i}_{k}\wedge\omega^{k}_{j}. \tag{S8}
\end{align*}
For a dislocation defect, the curvature vanishes while the torsion is singular, $T^i=b^i \delta({\mathbf l})$ with ${\mathbf b}$ as the Burgers vector and ${\bf l}$ as the position of the dislocation line. Since curvature tensor vanishes for a dislocation, the corresponding spin-connection can be set to zero.

\subsection*{B.1. Edge dislocations}

In case of an edge dislocation oriented along the $\hat{z}$-direction, the vielbein takes the form
\begin{equation*}
{\hat E}\equiv E^{i}_{\alpha}=
\begin{pmatrix}
1-\frac{by}{2\pi r^2}&\frac{bx}{2\pi r^2} &0\\
0 &1&0\\
0&0&1\tag{S8}
\end{pmatrix},
\end{equation*}
where $b$ is the magnitude of the Burgers vector $\mathbf{b}$, which is assumed to be along the $\hat{x}$-direction $\mathbf{b}=b\mathbf{e}_{x}$, and $r^2\equiv x^2+y^2$. The inverse of the above vielbein is readily found to be
\begin{equation*}
{\hat E}^{-1}\equiv E^{\alpha}_{i}=
\begin{pmatrix}
\left(1-\frac{by}{2\pi r^2}\right)^{-1}&-\frac{bx}{2\pi r^2}\left(1-\frac{by}{2\pi r^2}\right)^{-1} &0\\
0&1&0\\
0&0&1\tag{S9}
\end{pmatrix}.
\end{equation*}
Since for an elementary dislocation $b=a$ with $a$ as the lattice constant,  the distortion field ${\bf E}_i={\bf e}_i+{\bm \varepsilon}_i$ to the leading order in $a/r$ is then readily obtained
\begin{equation}\nonumber
{\bm \varepsilon}_x=\frac{ay}{2\pi r^2}{\bf e}_x=\frac{y}{2\pi r^2}{\mathbf b}\qquad {\bm \varepsilon}_y=-\frac{ax}{2\pi r^2}{\bf e}_x=-\frac{x}{2\pi r^2}{\bf b}.
\end{equation}
The corresponding gauge potential ${\bf A}=-{\bm \varepsilon}_i\cdot {\bf K}_{\rm inv}$, with ${\bf K}_{\rm inv}$ as the band-inversion momentum, can be then straightforwardly obtained and is given by Eq.\ (2) in the main text.

 We can now consider the effect of an edge dislocation in a 3D topological insulator. We note that along the core of the dislocation translational symmetry is preserved. Henceforth, $k_{z}$ is a good quantum number and we obtain a family of two-dimensional Hamiltonians
\begin{equation*}
H(x,y,z)=\sum_{k_{z}}e^{ik_{z}z} H_{eff}(x,y, k_{z}).
\end{equation*}
The dimensionally-reduced Hamiltonian $H_{eff}(x,y, k_{z}))$ can then be treated using well-known methods [S2,S4]. Hence, we can determine the spectrum of dislocation modes directly.

We now make this more concrete for the model with the Hamiltonian \eqref{eq::tb}. We assume that the system is in the  $T\text{-}p3(4)_{R}$ or $R$ phase [S6] as shown in Fig. S2. Using $\mathbf{k}\rightarrow\mathbf{k}+\mathbf{\mathbf{A}}=\mathbf{k}+\frac{1}{2r}\mathbf{e}_{\varphi}$, straightforward calculation yields
\begin{equation*}\label{eq::2D}
H_{eff}(x,y, k_{z}))=i\gamma_{r}\partial_{r}+i\gamma_{\varphi}(\frac{\partial_{r}}{r}+\frac{1}{2r})+\gamma_{0}[{M}-2B(\triangle+
\frac{i}{r^{2}}\partial{\varphi}-\frac{1}{4r^{2}})]+\sin(k_{z})\gamma_{3}, \tag{S10}
\end{equation*}
where
\begin{equation*}\label{eq::Mass}
{M}=m-8B-2B(1-\cos k_{z})=\hat{M}-2B(1-\cos k_{z} ).
\end{equation*}
 Here, the Laplacian $\triangle=\partial_{r}^2+\frac{\partial_{r}}{r}+\frac{1}{r^2}\partial_{\varphi}^{2}$, and $(r,\varphi)$ are the usual polar coordinates related to the Cartesian ones by $x=r\cos\varphi,\,\,\, y=r\sin\varphi$.
 To analyze the propagating midgap modes bound to the dislocation, let us first neglect the last term in the Hamiltonian $\sim \sin k_z$. In the remaining Hamiltonian the term  ${M}$ plays the role of the effective mass in the two-dimensional BHZ model. Therefore, this Hamiltonian hosts zero modes for the values of the parameters $m, B, k_z$ obeying $0<{M}/B<4$. Now, take $k_z=\pi$.
Then, as $8<m/B<12$, it follows that $4<\hat{M}/B<8$, which is precisely the condition for a zero mode solution bound to the dislocation. The solution of Eq.\ \eqref{eq::2D} without the last term $\sim \gamma_3$ has been derived and discussed in detail in Refs. [S2,S4]. More specifically, in this parameter range the system entails the $T\text{-}p4$ or $M$ phase in the reduced 2D Hamiltonian. Similarly, the next $k_{z}$ value in a finite system results in a dislocation mode if the effective mass $\hat{M}(k_{z})$ still satisfies the topological condition $4<\hat{M}/B<8$. Moreover, we can exactly determine the corresponding spectrum by reintroducing the last term in the Hamiltonian (\ref{eq::2D}), $\Delta_3\equiv\sin(k_{z})\gamma_{3}$. Since $k_z$ commutes with the Hamiltonian and $\gamma_3$ anticommutes with the other $\gamma$-matrices in the Hamiltonian, the term $\Delta_3$ acts as a mass term for the modes comprising zero-energy subspace in its absence. The corresponding gap is thus $\pm|\sin(k_{z})|$ with respect to the zero energy, and in the thermodynamic limit these midgap modes are linearly dispersing along the dislocation line with the velocity set by the hopping in this direction, in agreement with the $\mathbf{K}\text{-}\mathbf{b}\text{-}{\mathbf t}$ rule. These conclusions are schematically shown in Fig. S2. By displaying the allowed values of $k_z$ in a finite system, $k_{z}=2\pi n/L$, with $L\in {\mathbb Z}$ and $n=0,1,2...L-1$, on the unit circle in the imaginary plane, the real part characterizes the effective $\hat{M}/B$, while the imaginary part signals the anticipated energy of the dislocation mode. We note that the shift in $\hat{M}$ indeed ensures that  the $T\text{-}p3(4)_{R}$ phase hosts no dislocation modes with $k_{z}=0$ and in the thermodynamic limit all the propagating modes indeed descend from the zero modes at the band-inversion momentum $(\pi,\pi,\pi)$, as expected from the $\mathbf{K}\text{-}\mathbf{b}\text{-}{\mathbf t}$ rule.

\begin{figure*}[h]
 \includegraphics[scale=0.8432]{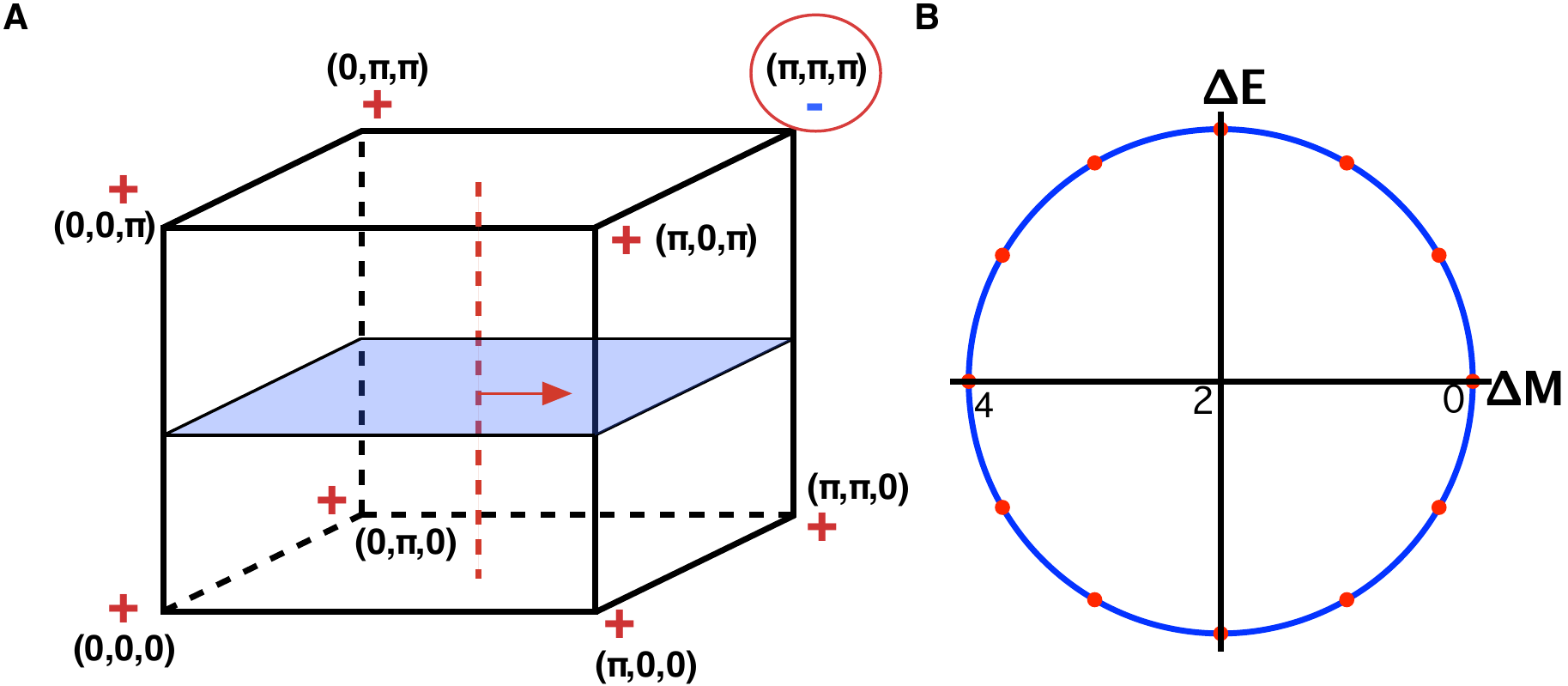}
\caption{\label{Fig. S2}Figure\ S2: The left panel shows the schematic configuration of the $T\text{-}p3(4)_{R}$ or $R$ phase in the Brillouin zone with an edge dislocation with Burgers vector $\mathbf{b}=\mathbf{e}_{x}$ and is oriented along the $\hat{z}$ direction in the real space. The right panel displays the modes $k_{z}$ on the unit circle in the complex plane. The imaginary part then encodes for the mass gap of the dislocation mode with respect to the zero mode, whereas the real part relates to the shift $\Delta M(k_{z})= -{M}/B+\hat{M}/B=2(1-\cos(k_{z}))$  of the effective mass parameter in the associated 2D BHZ model. }
\end{figure*}

The above analytical results can easily be verified by numerical computations, and the two show an excellent agreement. In Fig. S3 we show the spectrum of a $12\times12\times12$ system on a torus (periodic boundary conditions) and the real space localization of the dislocation modes. We note that the modes are neatly localized and come as two Kramers pairs, one from the dislocation and one from the anti-dislocation. Moreover, since $\exp(i k_{z})$ takes values in the set comprising  of the  twelfth roots of unity,  we can compare the energy gaps with the anticipated $\sin(k_{z})$ dependence and find agreement up to two decimals. In addition, we change the mass parameter $m$ to confirm that modes can be added or removed from the spectrum consistent with the condition in Eq.\ \eqref{eq::2D}, see Fig. S4. These results  reproduce for various sizes of the system (ranging up to 16 sites in length), confirming the presence of the zero mode together with the descendant propagating states, as predicted by the $\mathbf{K}\text{-}\mathbf{b}\text{-}{\mathbf t}$ rule. Furthermore, when the next-nearest neighbor hoppings are included, the phase $T-p4_M$ protected by both TRS and crystal symmetries with the band inversions at the momenta $(0,\pi,\pi),(\pi,0,\pi)$, and $(\pi,\pi,0)$ can be realized, see Fig.\ S1. In that case, a dislocation with Burgers vector ${\mathbf b}=a {\bf e}_x$ and oriented along the $z$ axis, $\hat{\bf t}={\bf e}_z$, in both planes $k_z=0$ and $k_z=\pi$ acts as $\pi$-flux, and thus a double Kramers pair is expected. This is indeed what we find numerically, see Fig. S5.

\begin{figure*}[h]
 \includegraphics[scale=0.8432]{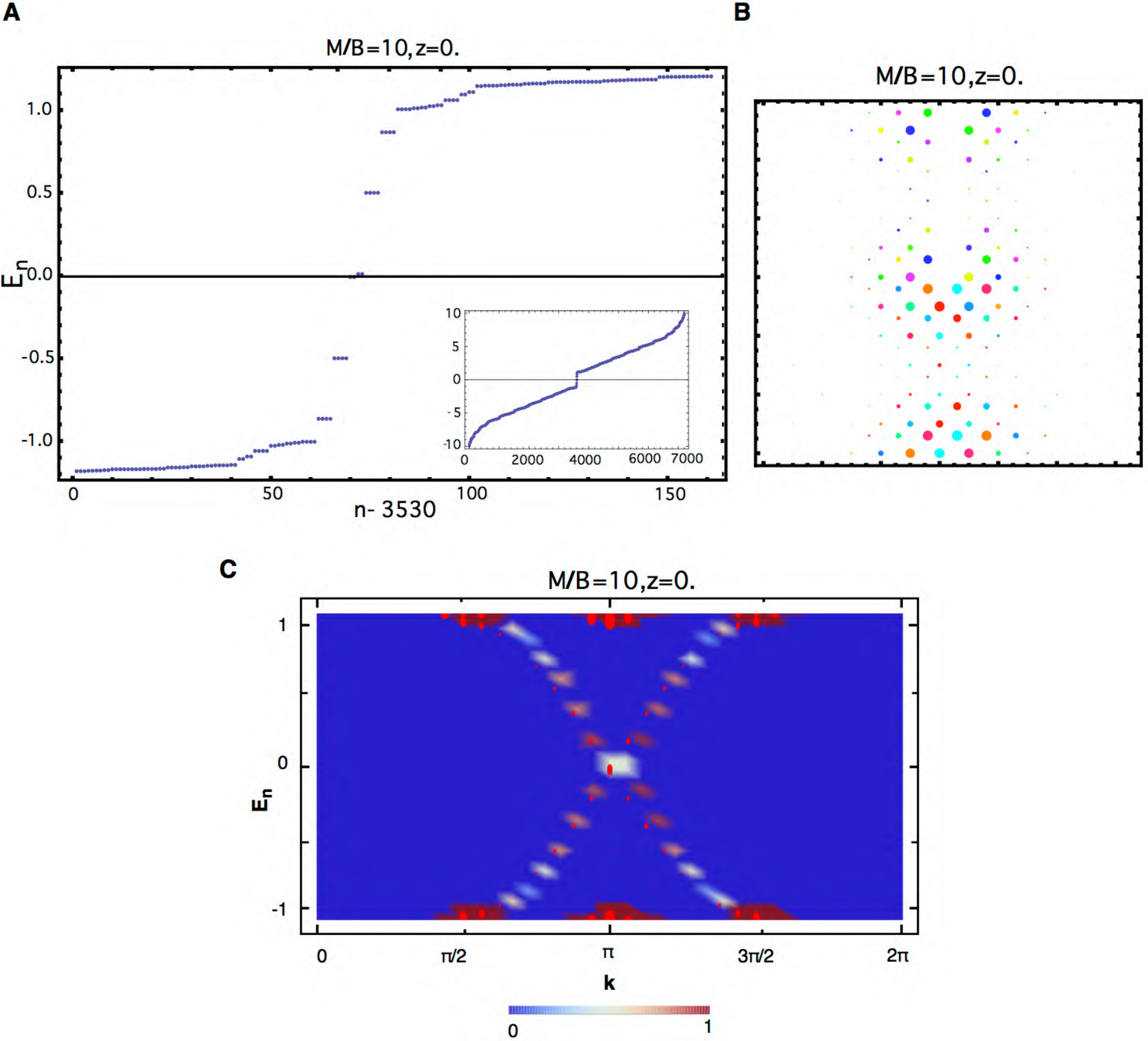}
\caption{\label{Fig. S3}Figure\ S3: The spectrum of the $12\times12\times12$ system (${M}/B=10$) in the $T\text{-}p3(4)_{R}$ phase with the edge dislocation with periodic boundary conditions. The dislocation modes come in degenerate  Kramers pairs, originating from the dislocation and anti-dislocation (panel A). The inset displays the complete spectrum. We note that the energy levels of these modes in the gap are located at $E_{n}=0$,  $\pm E_{N}=0.50$ and at $\pm E_{N}=0.86$, in agreement with the theory. Panel B shows the real space localization of the dislocation modes for a fixed ${\bf e}_z$ plane. The weight of the wavefunction is represented here by the radius of the circles, whereas the color indicates the phase following the same conventions as in Fig 2 in the main text.  Due to the translational symmetry, the ${\bf e}_z$ planes  are identical. Finally, panel C shows the the spectral density as function of the momentum ${k}={k}_{z}$ along the dislocation line for a $12\times12\times28$ system. This spectral density is also plotted as circles, where the radius indicates the weight, in order to further emphasize the excellent agreement with numerics. }
  \end{figure*}

\begin{figure*}[h]
\begin{center}
 \includegraphics[scale=0.8432]{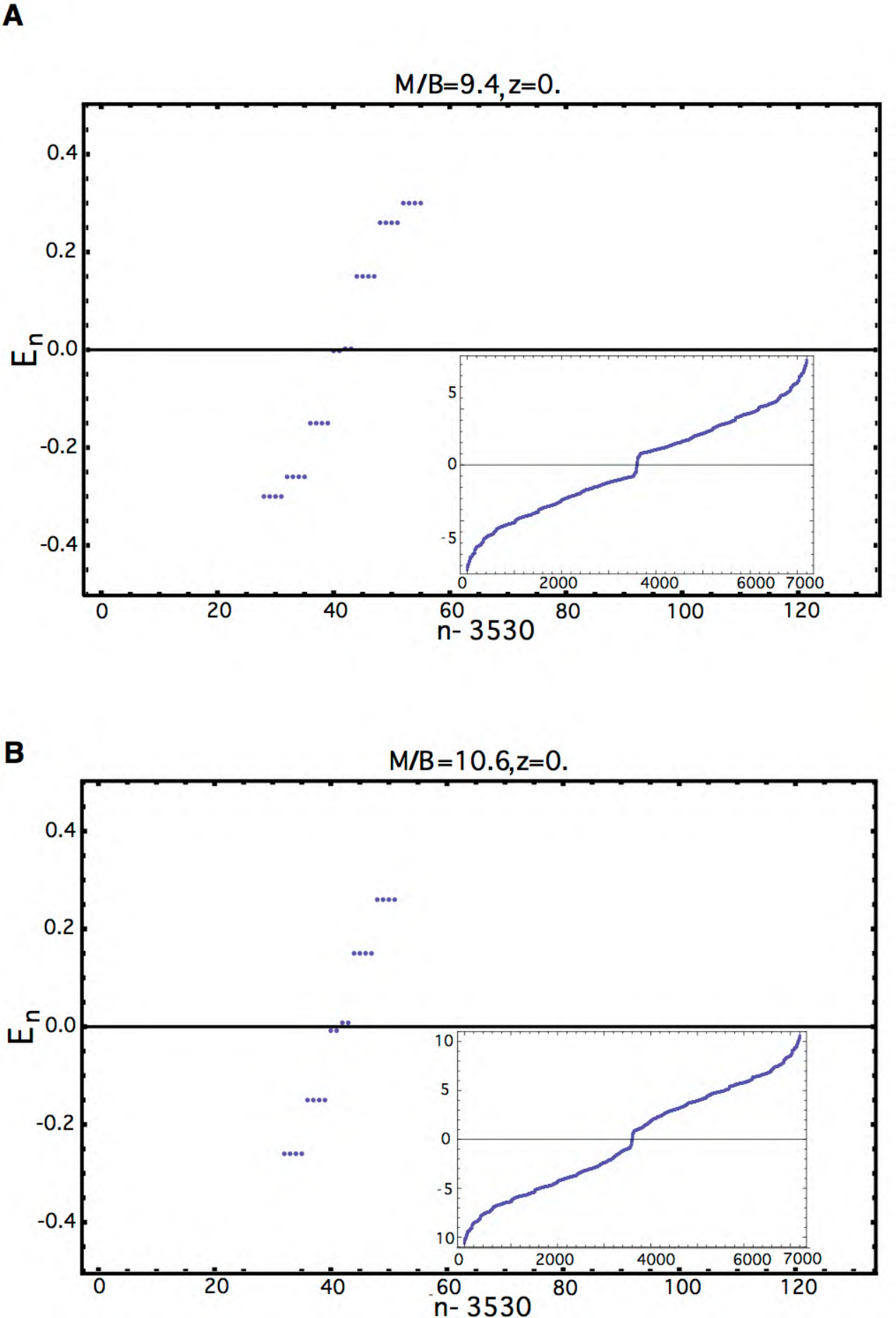}
\caption{\label{Fig. S4}Figure\ S4: The spectra for two $12\times12\times12$ systems on a torus, with ${M}/B=9.4$ and ${M}/B=10.6$ in the presence of an edge dislocation with Burgers vector ${\bf b}= {\bf e}_x$. In this instance $A_{z}$ was taken to be 0.3, shifting the energy levels to assure that they lie in the gap. The energy levels again take values that match the evaluation outlined above. Moreover, we observe that
the ${M}/B=9.4$ has one extra level, which is consistent with the mass condition  $4<\hat{M}<8$.}
 \end{center}
 \end{figure*}

 \begin{figure*}[h]
\begin{center}
 \includegraphics[scale=0.8]{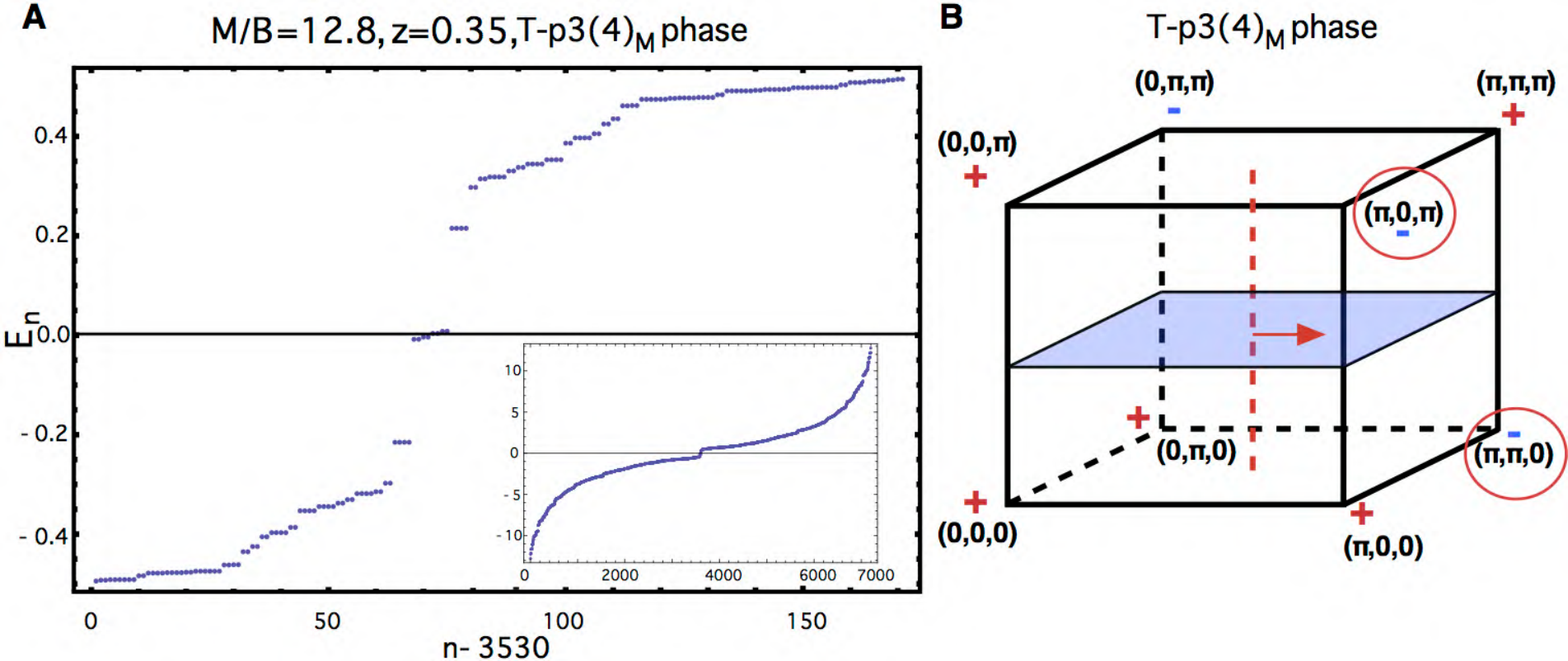}
\caption{\label{Fig. S7}Figure\ S5: Effect of an edge dislocation in the $T\text{-}p3(4)_{M}$ phase. The right panel shows the topological configuration, with an edge dislocation oriented along the $\hat{z}$ direction and with $\mathbf{b}=\mathbf{e}_{x}$. The left shows the spectrum,  and as expected both $k_{z}=0$ and $k_z=\pi$ planes host a $\pi$ flux yielding two Kramers pairs of zero modes. The resultant energy levels of the descendant states at finite energy are then fourfold degenerate. The modes originating from the  $k_{z}=\pi$ plane result from the corresponding two-dimensional $p4$ phase, whereas the $k_{z}=0$ plane is in the  $T\text{-}p4$ phase.}
 \end{center}
 \end{figure*}

It is straightforward to generalize the explained reasoning to other phases. Let us illustrate this with  the primitive tetragonal system taking $A_{x}=A_{y}=\frac{1}{2}A_{z}$ and $B_{x}=B_{y}=\frac{1}{2}B_{z}$ in Eq.\ (\ref{eq:tetragonal}). For these parameters we obtain the weak $p4_{M,R}$ phase with band-inversions at $M$ and $R$ points. This phase is  displayed in Fig. S6, and is not protected by either time-reversal symmetry or 3D space group symmetry, as opposed to a valley phase. By inserting an edge dislocation with ${\mathbf b}={\bf e}_x$ and $\hat{\mathbf t}={\bf e}_z$ we see that both time-reversal invariant planes orthogonal to $\hat{\mathbf t}$ are  topologically nontrivial and thus yield zero modes, which in turn produce descendant propagating modes along the dislocation.
We find the propagating modes numerically
 for the systems with $10\times10\times10$ and a $12\times12\times12$ sites, see Fig. S6, and with the plot of the spectral weight of the dislocation modes shown in Fig.\ S7. Notice the doubling of the Dirac cone since in this phase there are two nontrivial topologically nontrivial planes orthogonal to the dislocation line each yielding a $\pi$ flux. In turn, this nontrivial flux produces a double Kramers pair of the zero modes, which yields descendent propagating states along the dislocation defect.

\begin{figure*}[h]
\begin{center}
 \includegraphics[scale=0.8]{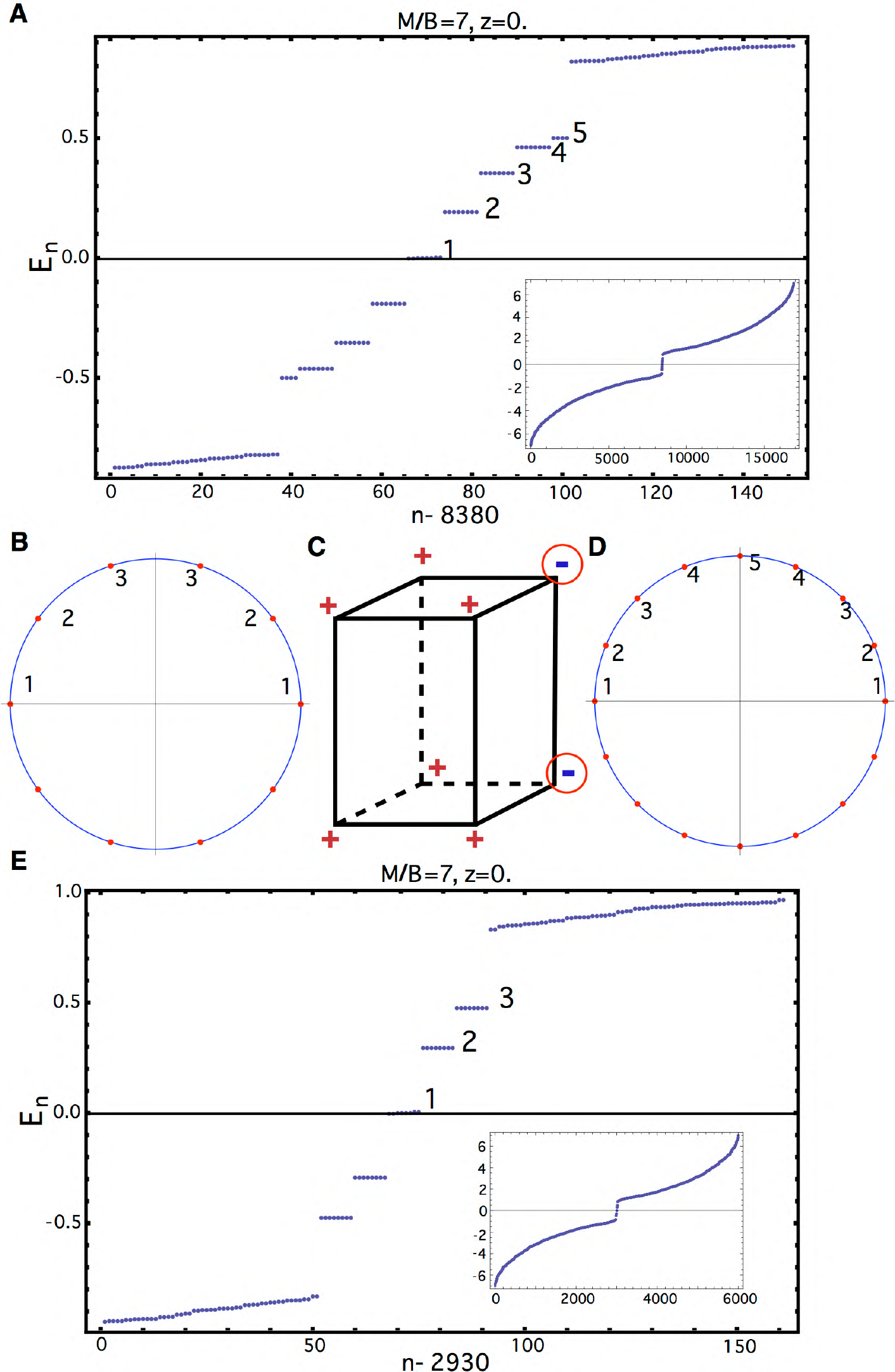}
\caption{\label{Fig. S5}Figure\ S6: The spectra of the tetragonal system with $12\times12\times12$ (A and E) and a $10\times10\times10$ (D and E) sites on a torus in the $p4_{M,R}$  phase, with the topological configuration shown in panel C, and an edge dislocation with ${\bf b}={\bf e}_x$. We note the eightfold degeneracy per level, consistent with the $\mathbf{K}\text{-}\mathbf{b}\text{-}{\mathbf t}$ rule. Namely as the $k_{z}=0$ and $k_{z}=\pi$ planes host the effective $\pi$ fluxes, the modes denoted by 1 in panels A and E are indeed localized  around the defect as anticipated. Modes 2-5 and 2-3 in the same panels descend from these zero-energy modes with the spectra matching the analytical results.}
 \end{center}
 \end{figure*}

\begin{figure*}[h]
\begin{center}
 \includegraphics[scale=0.8]{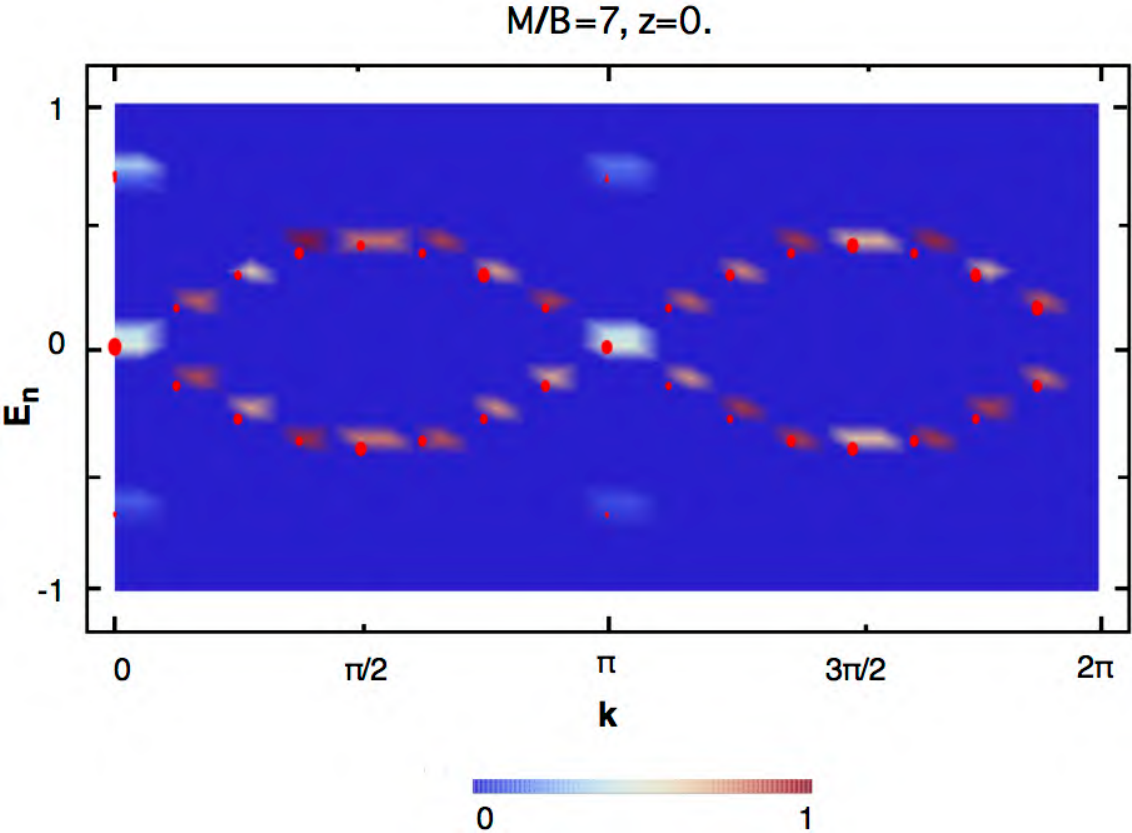}
\caption{\label{Fig. S6}Figure\ S7: The spectral density of the  $p4_{M,R}$  phase as function of the momentum ${k}=k_{z}$, in case of $16\times16\times16$ system with periodic boundary conditions. We observe a double Dirac cone situated at $k_{z}=0$ and $k_{z}=\pi$, consistent with the findings in Fig.\ S6. Conventions are identical as in Fig.\ S3.}
 \end{center}
 \end{figure*}

\clearpage
\newpage
 \subsection*{B.2. Screw dislocations}
In three dimensions the Burgers vector can also be oriented parallel to the dislocation line. In contrast to the edge dislocation, which is essentially a 2D defect pulled out in the third dimension, the resulting screw dislocation is an intrinsic 3D defect. For a screw dislocation with the Burgers vector oriented along the $\hat{z}$-axis, ${\mathbf b}=b{\bf e}_z$, the vielbeins are given by
\begin{equation*}
{\hat E}\equiv E^{i}_{\alpha}=
\begin{pmatrix}
1&0&0\\
0&1&0\\
-\frac{by}{2\pi r^2} &\frac{bx}{2\pi r^2}&1\\
\end{pmatrix},
\end{equation*}
and
\begin{equation*}
{\hat E}^{-1}\equiv E^{\alpha}_{i}=
\begin{pmatrix}
1&0&0\\
0&1&0\\
\frac{by}{2\pi r^2} &-\frac{b x}{2\pi r^2} &1\\
\end{pmatrix}.
\end{equation*}
The corresponding distortion field, taking into account that for an elementary dislocation $b=a$, is then given by
\begin{equation}
{\bm \varepsilon}_x=\frac{ay}{2\pi r^2}{\bf e}_z=\frac{y}{2\pi r^2}{\mathbf b}\qquad {\bm \varepsilon}_y=-\frac{ax}{2\pi r^2}{\bf e}_x=-\frac{x}{2\pi r^2}{\bf b},
\end{equation}
and the gauge potential is given by Eq.\ (2) in the main text.

We note for a screw dislocation, the gauge potential can only be finite if $K_{{\rm inv},z}\neq0$. In the case of a screw dislocation, $k_{z}$ is still a good quantum number, and henceforth we find the same equation for the zero-energy modes and the corresponding descendant states \eqref{eq::2D}. In Figs.\ S8 and S9, we present the numerical analysis of the spectrum of the Hamiltonian in the $T-p3(4)_R$ phase in the presence of a screw dislocation. We obtain in essence the same spectrum as for the edge dislocation.
For various system sizes, we find precisely the number of dislocation modes in the spectrum according to the topological condition below Eq. \eqref{eq::2D}. Also, we find that the energy difference between the levels is proportional with the $A_{z}$ coefficient. We do find however that the energy levels have shifted as compared with the ones obtained in the presence of an edge dislocation in the same phase. Tuning the mass parameter ${M}/B$ deeper into the phase, (closer to the value of this parameter $M/B=12$) results in levels closer to the analytically calculated value, as expected from the finite-size effects, and the fact that the matrix elements are now twisted with a factor $e^{ik_{z}}$ along the dislocation direction. We observe in addition that, in general, the form of the underlying metric $ds^2=dr^2+r^2d\vartheta^2+(dz+\beta d\vartheta)^2$ may be mapped with $z\rightarrow z+\beta\theta$ to a flat space with nontrivial quasi-periodic boundary conditions, which indeed result in energy shifts, which we do not explicitly compute here but this in principle can be done using the outlined procedure.

In the translationally active $T\text{-}p3(4)_M$ phase, obtained from the Hamiltonian \eqref{eq::tb} and for the values of the parameters as shown in Fig.\ S1, a screw dislocation with the Burgers vector ${\bf b}={\bf e}_x$ produces $\pi$ fluxes in the plane $k_x=\pi$, which hosts a valley phase. Therefore, according to the  ${\bf K}\text{-}{\bf b}\text{-}{\bf t}$ rule, we expect two Kramers pairs of the dislocation modes originating from the band-inversion at the $M$ and the $X'$ points. Our numerical computations indeed confirm this prediction, as shown in Fig.\ S10.
\begin{figure*}[h]
\begin{center}
 \includegraphics[scale=0.73]{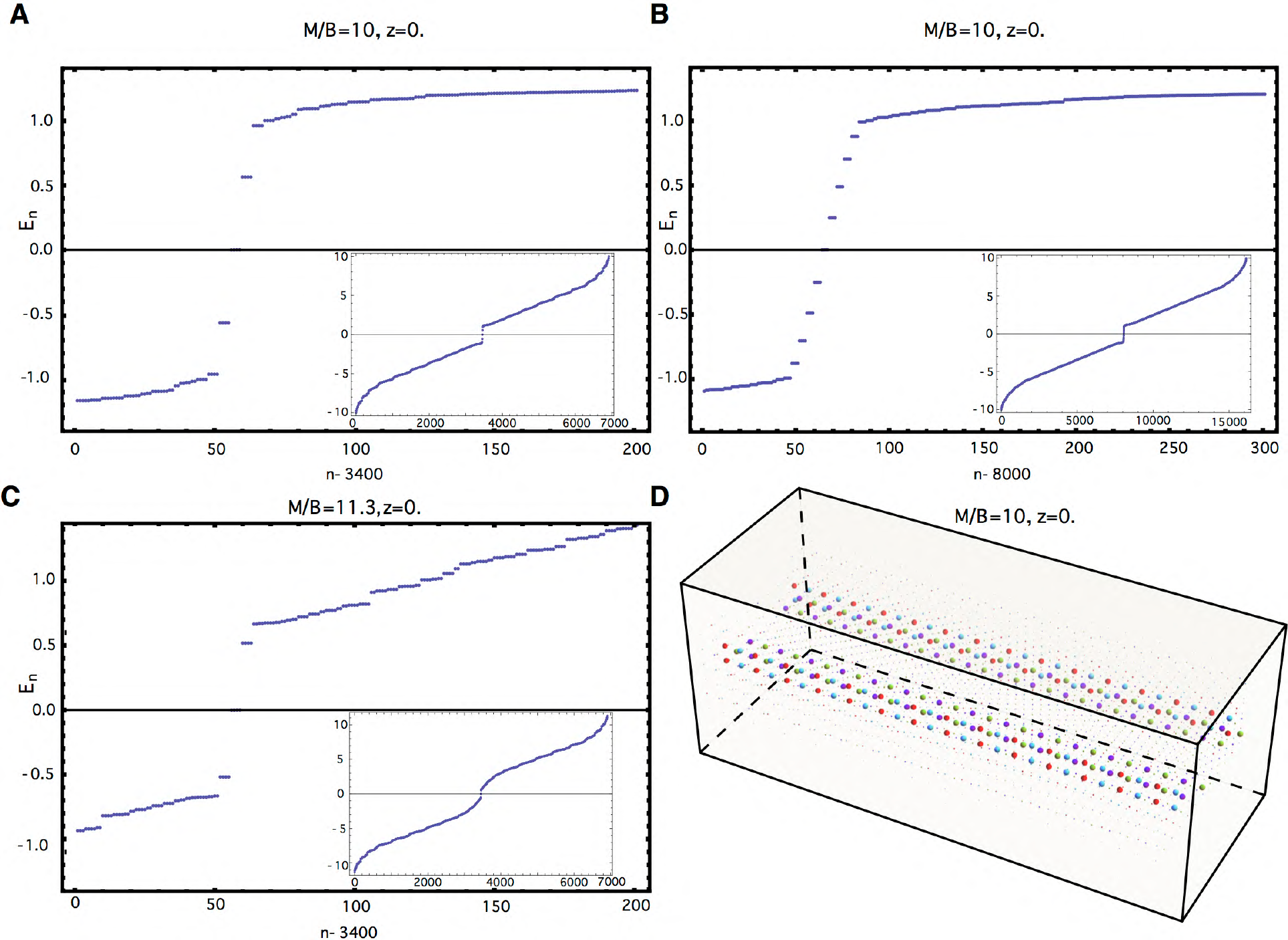}
\caption{\label{Fig. S8}Figure\ S8: Numerical results concerning the effect of a screw dislocation, with $\mathbf{b}=\mathbf{e}_{x}$ oriented in the ${\bf e}_{x}$ direction. Panel A shows the spectrum of the modes on a $12\times12\times12$ system with periodic boundary conditions. The modes are not exactly at $\pm$ 0.50 as expected from the continuum model. The deviation of the energy levels from the anticipated value as obtained from the analytical treatment becomes smaller for larger systems sizes. Panel B shows a $28\times12\times12$ system showing that the splitting of the energy levels converges to the result from the continuum model. Similarly the agreement with numerical results is also better in the $T-p4(3)_{R}$ phase when the system is closer to the transition point ${M}/B=12$ (C). Finally panel D shows the real space localization of the mode with the color coding  shown above.
}
 \end{center}
 \end{figure*}

\begin{figure*}[h]
\begin{center}
 \includegraphics[scale=0.73]{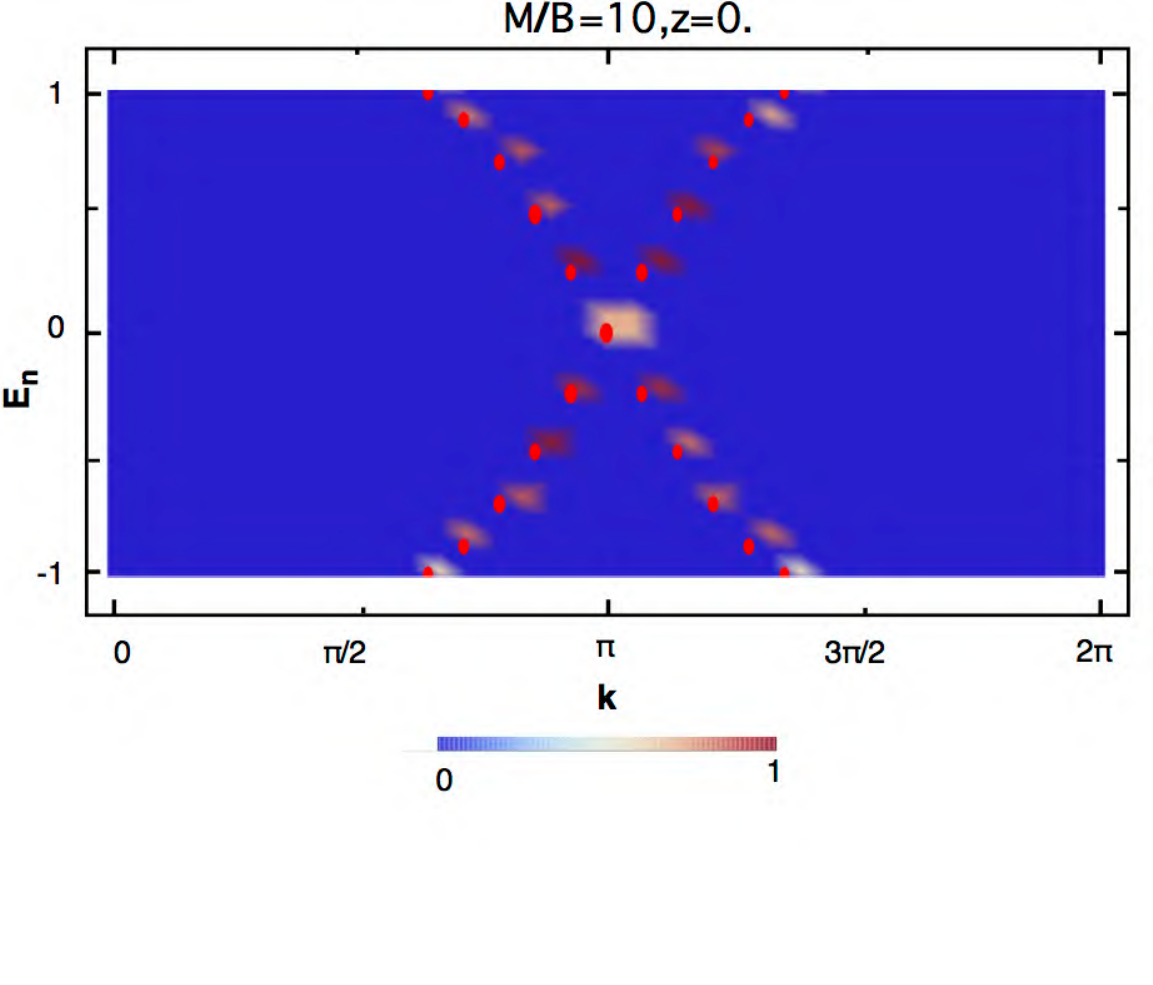}
\caption{\label{Fig. S9}Figure\ S9: The spectral density of a screw dislocation, with Burgers vector $\mathbf{e}_{x}$, in the $T\text{-}p3(4)_{R}$  phase of a $28\times12\times12$ system with periodic boundary conditions. The presentation is similar to the above Figures. The horizontal axis shows the momentum ${k}=k_{x}$ along the dislocation line. Although the energy levels are shifted due to the finite size of the system, the spectrum displays again a zero mode at $k_{x}=\pi$ and the expected number of descendant states that form a Dirac cone, analogous to the case of an edge dislocation.}
 \end{center}
 \end{figure*}

 \begin{figure*}[h]
\begin{center}
 \includegraphics[scale=0.73]{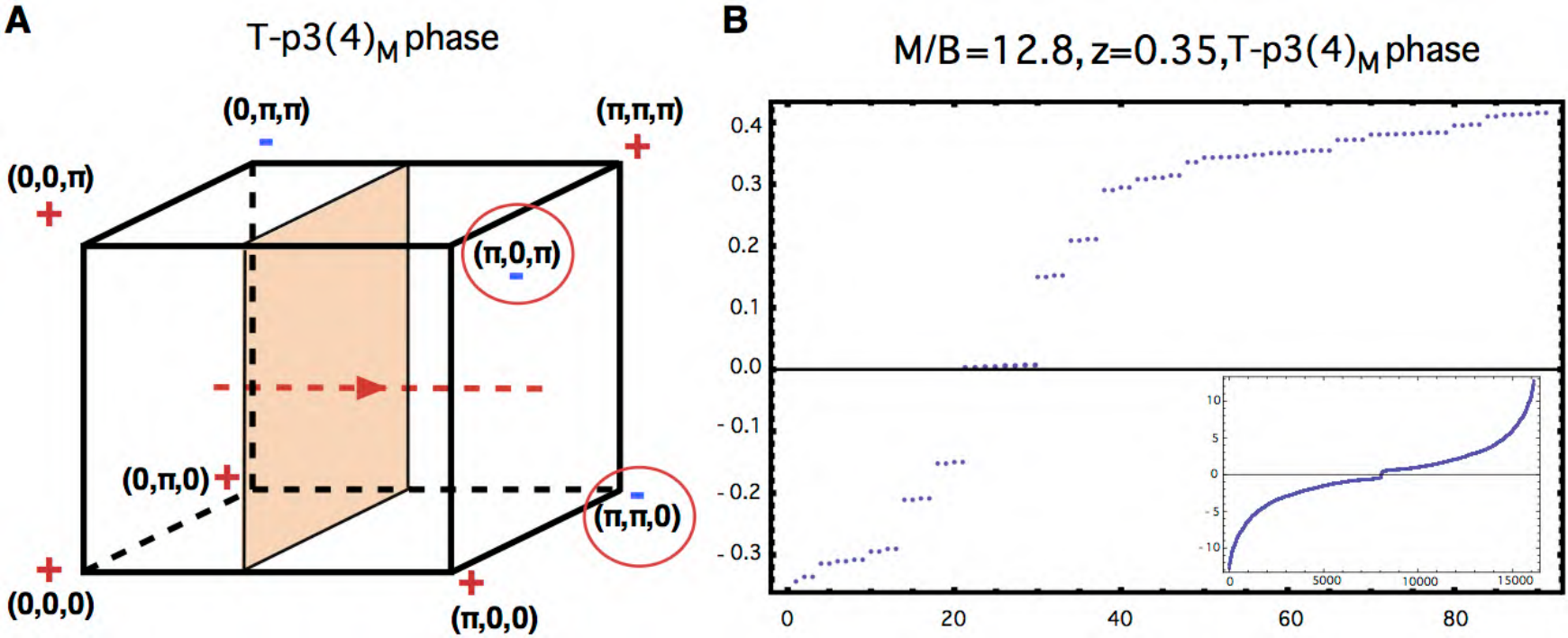}
\caption{\label{Fig. S10}Figure\ S10: Numerical results concerning the effect of a screw dislocation in the  $T\text{-}p3(4)_{M}$  phase. (A) The topological configuration of the $T\text{-}p3(4)_{M}$  phase.
The screw dislocation, with Burgers vector ${\mathbf b} = {\bf e}_x$,  acts on the encircled TRI momenta in the $\hat{y}-\hat{z}$ planes. (B) The resulting spectrum of a $28\times12\times12$ system with periodic boundary condition in the $T\text{-}p3(4)_{M}$  phase. The resulting spectrum shows a double pair of 'parent' helical zero modes. These modes originate from the encircled momenta, which form a 2D valley phase in the plane $k_x=\pi$, since these momenta are related by a threefold rotation around the axis connecting the $\Gamma$ and the $R$ points.   }
 \end{center}
 \end{figure*}

Let us now consider a dislocation loop, which can be thought of as a connected channel of screw and edge dislocations. Although the Burgers vector $\mathbf{b}$ of any dislocation remains constant [S7], the planes orthogonal to the dislocation line are different for the edge and screw dislocation part of the circuit.  In general one can consider the full scattering problem of an edge and screw dislocation [S8,S9], which essentially pertains to matching the phases of the solutions of Eq.\ \eqref{eq::2D}.
 Due to time-reversal symmetry, however, the helical modes do not backscatter, and we can thus conclude that dislocation loops have helical modes along the core when both the edge and the screw dislocation possess the propagating modes. We note that this implies that the existence of the dislocation modes in the loop is also dictated by the ${\bf K}\text{-}{\bf b}\text{-}{\bf t}$ rule.
Accordingly, the $T\text{-}p3(4)_{R}$ phase with a dislocation loop, see Fig S2., has dislocation modes along the loop for any orientation of the Burgers vector $\mathbf{b}$. In contrast, inserting a dislocation loop in the $\hat{x}-\hat{z}$ plane, with  $\mathbf{b}=\mathbf{e}_{x}$, in the $p4_{M,R}$ phase shown in Fig. S6 only binds dislocation modes to the edge dislocation part of the circuit, resulting in the exact same spectrum shown in Fig. S6. Changing the orientation of the Burgers vector to $\mathbf{e}_{z}$ then does result in dislocation modes along the loop, as the gauge potential ${\bf A}$ is nontrivial in all the planes normal to the dislocation lines and these planes are topologically nontrivial.
At last, changing the Burgers vector to the $\mathbf{e}_{y}$ direction, it is evident that the effective systems in each plane either has no flux or acquires $2\pi$-flux. As a result we anticipate no modes, analogously to the case of a screw dislocation with ${\bf b}={\bf e}_x$. The latter is confirmed by numerical computations, see Fig. S11C.

Finally, we address the full compatibility of the outlined analysis with the underlying characterization of topological insulators with different space groups in terms of the band-inversions.  Consider, for example,  the composite phase $T\text{-}pm\bar{3}m\oplus T\text{-}4p3_{X}$, which is equivalent to the $T\text{-}4p3_{M}\oplus T\text{-}4p3_{R}$ or $\Gamma\oplus XYZ=MX'Y'\oplus R$ phase with the dislocation loop in the $y-z$ plane, as previously described, see Fig. S11. We expect here the response of the dislocations to be the same with both choices of the band-inversions. We see that applying the ${\bf K}\text{-}{\bf b}\text{-}{\bf t}$ rule to either set of the band inversions produces the same outcome in terms of the fluxes acting on the planes.
Along the edge dislocation, the $k_{z}=0$ plane hosts a $\pi$ flux, in contrast to the $k_{z}=\pi$ plane. Similarly, for the screw dislocation part, which has momentum in the $k_{x}$ direction, we see that the $k_{x}=\pi$ plane hosts a zero mode. Note that in the latter case, the screw dislocation acts as a $\pi$-flux in the phase, which with respect to the $k_x=\pi$ plane, may be thought of as the  $T\text{-}p4mm$ or $\Gamma$ phase, since the $(\pi,0,0)$ momentum acts as a $\Gamma$ point in this plane.  The descendant values of $k_{x}$ thus contribute dislocation modes to the spectrum as long as the effective ${M}/B$ parameter in the reduced model is in the range corresponding to the $T\text{-}p4mm$ in 2D. As result, we find that the complete dislocation loop has modes along the core.

\begin{figure*}[h]
\begin{center}
 \includegraphics[scale=0.73]{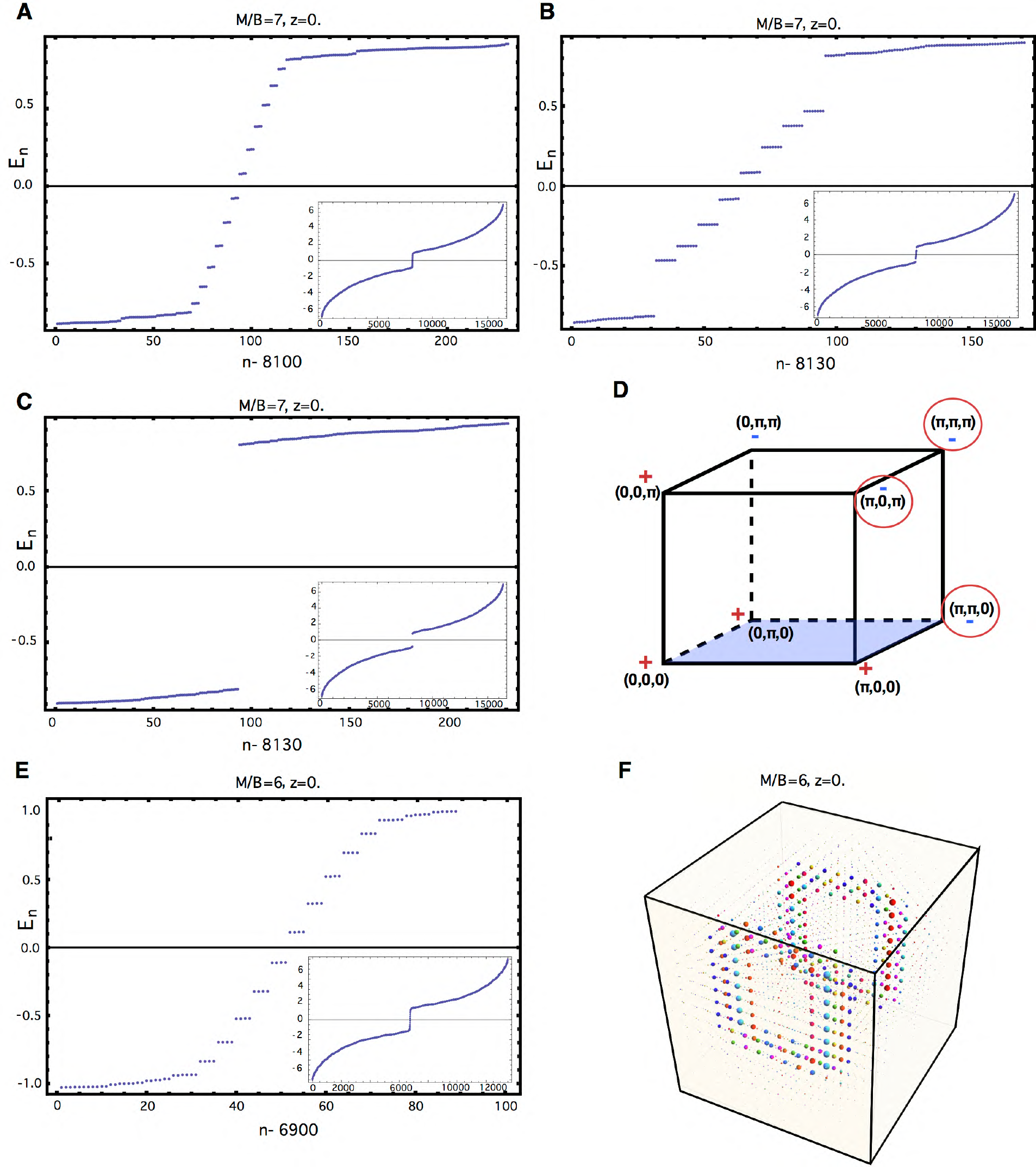}
\caption{\label{Fig. S11}Figure\ S11: Dislocation loop in the $\hat{x}-\hat{z}$ plane with Burgers vector $\mathbf{b}=\mathbf{e}_{x}$. Panels A to C show the spectra  corresponding to the dislocation loops in Fig. 1 in the main text.  Namely, considering systems with periodic boundary conditions, we tune the system to the respective phases and get the anticipated dislocation modes in the spectrum. Panel D indicates the electronic topological configuration of the $T\text{-}4p3_{M}\oplus T\text{-}4p3_{R}$ phase.  As each plane hosts an effective $\pi$-flux problem, inserting the same dislocation loops gives the anticipated spectrum (E), the modes of which show the familiar real space localization (F).  }
 \end{center}
 \end{figure*}

\clearpage
\newpage

Supporting References:

[S1] \textrm{Qi, X.-L.} \& \textrm{Zhang, S. C.}, \textrm{Topological insulators and superconductors}. {\it Reviews of Modern Physics} {\bf 83},
 1057-1110 (2011).\\

\noindent
[S2] \textrm{Juri\v{c}i\'{c}, V.},  \textrm{Mesaros, A.}, \textrm{Slager, R.-J.} \& \textrm{Zaanen, J.}, \textrm{Universal Probes of Two-Dimensional Topological Insulators: Dislocation and $\pi$ Flux}.
 {\it Phys. Rev. Lett.\/} {\bf 108}, 106403 (2012).\\

\noindent
[S3] \textrm{Kleinert, H.},
\textrm{ Gauge Fields in Condensed Matter, Vol. II}.
{\it World Scientific, Singapore, 1989}.\\

\noindent
[S4] \textrm{Mesaros, A.}, \textrm{Slager, R.-J.},  \textrm{Zaanen, J.} \& \textrm{Juri\v{c}i\'{c}, V.},
 \textrm{Zero-energy states bound to a magnetic $\pi$-flux vortex in a two-dimensional topological insulator}.
 {\it Nucl. Phys. B.\/} {\bf 867}, 977-991 (2013).\\

 \noindent
[S5]\textrm{Moore, J. E.} \& \textrm{Belents, L.}, \textrm{Topological invariants of time-reversal-invariant band structures}. {\it Phys. Rev. B\/} {\bf 75},
121306 (2007).\\

 \noindent
[S6] \textrm{Slager, R.-J.}, \textrm{Mesaros, A.},\textrm{ Juri\v ci\' c, V.} \& \textrm{Zaanen, J.}, \textrm{The space group classification of topological band-insulators}.
 {\it Nature Phys. \/} {\bf 9},  98-102 (2013).\\

\noindent
[S7] \textrm{ Landau, L. D.},\& \textrm{ Lifshitz, E. M.}, \textrm{Theory of Elasticity (Pergamon Press, New York, 1981).}\\

\noindent
[S8] \textrm{Kawamura, K.}, \textrm{A new theory on scattering of electrons due to spiral dislocations}. \textrm{Z. Phys. B.}
{\bf 29}, 101 (1978).\\

\noindent
[S9] \textrm{Kawamura, K.}, \textrm{Scattering of a tight-binding electron off an edge dislocation}. \textrm{Z. Phys. B.}
{\bf48},  201 (1982).

\end{widetext}

  \end{document}